\documentclass[11pt]{article}
\usepackage{a4wide}
\usepackage[normalem]{ulem}
\usepackage{graphicx}
\usepackage{amsmath,amssymb}
\usepackage{color}
\usepackage{slashed}
\usepackage{cancel}
\usepackage[bookmarks=true,bookmarksnumbered=true,setpagesize=false]{hyperref}

\usepackage{comment}
\usepackage{cases}

\newcommand{\tx}{\text}

\newcommand{\nn}{\nonumber\\}

\newcommand{\be}{\begin{equation}}
\newcommand{\e}{\end{equation}}
\newcommand{\aln}[1]{\begin{align}#1\end{align}}

\begin{document}
\title{
\vspace{-2cm}
\vbox{
\baselineskip 14pt
\hfill \hbox{\normalsize }} 
\vskip 1cm
\bf \Large Multi-critical point principle as the origin of classical conformality and its generalizations
\vskip 0.5cm
}
\author{
Hikaru~Kawai$^{ab}$\thanks{E-mail: \tt hikarukawai@phys.ntu.edu.tw}~ and  
Kiyoharu Kawana$^c$\thanks{E-mail: \tt kawana@snu.ac.kr}
\bigskip\\
\normalsize
 \it
 $^{a}$ Department of Physics and Center for Theoretical Physics,
 \\
 \normalsize
\it 
  National Taiwan University, Taipei, Taiwan 106,\\
 \normalsize 
 \it 
 $^{b}$ Physics Division, National Center for Theoretical Sciences, Taipei 10617, Taiwan 
 \\
\normalsize 
\it  
$^{c}$ Center for Theoretical Physics, Department of Physics and Astronomy,\\
 \normalsize
\it  Seoul National University, Seoul 08826, Korea
\smallskip
}

\date{\today}

\maketitle

\begin{abstract}\noindent 
Multi-critical point principle (MPP) is one of the interesting theoretical possibilities that can explain the fine-tuning problems of the Universe. 
% 
%Although there are a lot of different ways to explain it, the simplest 
It simply claims that ``the coupling constants of a theory are tuned to one of the multi-critical points, where some of the extrema of the effective potential are degenerate.'' 
One of the simplest examples is the vanishing of the second derivative of the effective potential around a minimum. 
This corresponds to the so-called classical conformality, because it implies that 
the renormalized mass $m^2$ vanishes. 
%, which is implicitly assumed in dimensional transmutation mechanism such as the Coleman-Weinberg (CW) mechanism. 
%
%the renormalized mass $m^2$ apparently corresponds to a critical point at which the vacuum expectation value (around the origin) disappears when  $m^2\nearrow 0$.      
%
More generally, the form of the effective potential of a model depends on several coupling constants, and we should sweep them to find all the multi-critical points. 
In this paper, we study the multi-critical points of a general scalar field $\phi$ at one-loop level under the circumstance that the vacuum expectation values of the other fields are all zero. For simplicity, we also assume that the other fields are either massless or so heavy that they do not contribute to the low energy effective potential of $\phi$.
%
%only the $\phi$ sector has dimensionful parameters at tree level.  
This assumption makes our discussion very simple because the resultant one-loop effective potential is parametrized by only four effective couplings. %thanks to the simple behavior of the background-dependent masses $M_i^{}(\phi)\sim \phi$ except for $M_\phi^{}(\phi)$. 
% 
%Then, it is possible to perform model-independent calculations of the vacuum structure.       
%
%Of course our analysis does not cover whole the multi-critical points of one-loop effective potential, but
Although our analysis is not completely general because of the assumption, it still can be widely applicable to many models of the Coleman-Weinberg mechanism and its generalizations. 
% discussed in many literatures.  
%we already have a general expression of one-loop effective potential (see Eq.~(\ref{CW potential})), studying multi-critical points of it without any assumptions is too complicated and less useful from phenomenological point of view.    
%
%In this paper, we will simply assume that only the $\phi$ sector has dimensionful parameters at tree level, which
After classifying the multi-critical points at low-energy scales, we will briefly mention the possibility of criticalities at high-energy scales and their implications for cosmology.    
\end{abstract}
\newpage

%\normalsize
%____________________Higgs Inflation____________________
\section{Introduction}
For the last few decades, the Standard Model (SM) has been tested by various experiments and observations, but any significant clues to new physics have not been discovered yet. %\footnote{\red{Should we mention muon $g-2$ ?}} 
Moreover, the renormalization group (RG) analysis based on the observed SM parameters supports the hypothesis that the SM is not much altered up to the Planck scale ~\cite{Degrassi:2012ry,Buttazzo:2013uya,Bednyakov:2015sca,Hamada:2012bp,Holthausen:2011aa,Bezrukov:2012sa}, which strongly suggests a possibility of ``{\it desert scenario}" such that the SM might be directly connected to %regarded as an effective theory of 
an ultraviolet complete theory.  

On the other hand, it is also true that there are a lot of mysteries and problems in particle physics and cosmology:  
% 
%Why the electroweak (EW) scale is so tiny compared to the Planck scale $10^{18}$~GeV ? 
%  
For example, why is the the electroweak (EW) scale $v=246$~GeV so small compared to the Planck scale $10^{18}$~GeV ? 
%we have not yet understood the origin of electroweak (EW) scale $v=246$~GeV, which is hugely small compared to the Planck or string scale $10^{18}$~GeV. % at which people believe that there must exist an unified theory which includes quantum gravity.  
% 
Why is the vacuum energy of the present universe so tiny ? 
%Why $U(1)$ charges are quantized ?
Why are there huge mass hierarchies among the SM   fermions ?
Why are there three generations ? What is dark matter ? What is the origin of baryon asymmetry of the universe ? etc.      
%Further, the Majorana-mass scale for the right-handed neutrinos is unknown in the SM with the seesaw mechanism~\cite{Minkowski:1977sc,Yanagida:1979as,GellMann:1980vs,Glashow:1979nm,Mohapatra:1979ia}.
%
Though we can easily make an extended model of the SM to explain some phenomenological aspects such as dark matter or baryon asymmetry, other issues related to the naturalness seem to require more radical and fundamental approaches than the ordinary quantum field theory.
% since they are usually difficult to be explained by simply extending the SM. 
%
%Each of those questions is just one of the aspects of more fundamental question; ``{\it Why low energy effective theory is the SM that we observe today ? }",  
%summarized by just one phrase, ``{\it Why the Universe is explained by the SM we observe ?}" 
%
%This is the most fundamental and natural question, but
%and we have not yet obtained a concrete explanation for it.  
%
%Recently, this situation also has been motivating us to study the landscale and swapland of theory space in quantum gravity where the latter corresponds to low-energy effective theories that can not be consistent with quantum gravity such as string theory \cite{Vafa:2005ui,Ooguri:2006in,ArkaniHamed:2006dz,Banks:2010zn,Brennan:2017rbf,Obied:2018sgi,Harlow:2018tng,Palti:2019pca}.   
%Moreover, the recent observations in cosmology, including that of the cosmic microwave background (CMB), have established the existence of (cold) dark matter (DM).
%

%It motivates us to consider a new particle whose interactions with the SM particles are relatively weak.   
The various observations of the SM may already provide important hints for solving these naturalness problems.
%The various observations of the SM might be already indicating  important hints to tackle those naturalness problems.  
%
%Let's take a deeper look at the RG behavior of the Higgs potential.  
%
It is a well-known fact that the Higgs potential becomes negatively large below the Planck scale when the top-quark mass is larger than some value. 
%about $173$~GeV.   
%
More precisely, the critical value of the top-quark pole mass for the theoretical border between stability and instability of the SM vacuum is $m_{t,\tx{critical}}^{\rm pole}\simeq 171.4$~GeV \cite{Hamada:2014wna,Hamada:2014xka}, which is consistent at the 1.4\,$\sigma$ level with the latest combination of the experimental results $m_t^{\rm pole}=172.4\pm 0.7$ GeV~\cite{PDG2020}. 
%(or metastability) of the effective Higgs potential for the observed Higgs mass $\simeq$125\,GeV;
%See also Refs.~\cite{Degrassi:2012ry,Buttazzo:2013uya,Bednyakov:2015sca}.
%\footnote{With the current central value $m_H=125.1\pm 0.1$\,GeV~\cite{PDG2020}, the critical top mass becomes $m_{t,\tx{critical}}^{\rm pole}\simeq171.2$\,GeV~\cite{Bednyakov:2015sca}. }
%
Surprisingly, for the critical top mass, the vacuum expectation value of the Higgs field 
%the degenerate minimum of the Higgs potential 
is nearly equal to the Planck scale, and such a behavior of the potential has a lot of implications to high energy physics and cosmology.  
This interesting behavior of the Higgs potential can be understood as one of the manifestations of the multi-critical point principle (MPP)~\cite{Froggatt:1995rt,Froggatt:2001pa,Nielsen:2012pu,Kawai:2011rj,Kawai:2011qb,Kawai:2013wwa,Hamada:2014ofa,Hamada:2014xra,Hamada:2015dja,Hamada:2015ria,Kannike:2020qtw}. %(see also Refs.~\cite{Bennett:1988xi,Bennett:1993pj}) 
%that ``coupling constants that are relevant at low energies are tuned to a multi-critical point around which the vacuum structure drastically changes when they are varied.'' 
%
Actually there are several manners to represent the MPP. A simple one is: ``The coupling constants of a theory are tuned to one of the multi-critical points of the vacua.  
In other words, some of the extrema of the effective potential should be degenerate." 
%Here, we should emphasize that we also include the local extrema in the meaning of ``the vacuum structure".   
%
In this sense, the MPP provides a natural fine-tuning mechanism for coupling constants in quantum field theory. %, and it has a great potential to explain the above fundamental questions. 
Note that the effective potential exists independently of the renormalization scale, up to the wave function renormalization of the field, for a given set of bare parameters. 
%
%In other words, 
In general, when there is only one mass scale involved, we can approximate the effective potential by lower order perturbations by choosing the renormalisation scale around that mass scale. 
In this paper, we discuss that non-trivial critical points exist near the mass scale where the running quartic coupling becomes zero. 

Here we give a short introduction to the MPP to clarify and visualize how the renormalized parameters are fixed by this principle.  
%
%Readers who are familiar with the MPP can skip this part. 
%
For simplicity, let us consider the effective potential $V(\phi)$ of a general scalar field $\phi$.   
In general $V(\phi)$ depends on the coupling constants $\{m^2,~\lambda,~g,~y,\cdots \}$ of the theory, and its extrema change as we vary them.  
Then, we can find critical points in the parameter space at which some of the extrema become degenerate.  
%
%The simplest example is $V(\phi)=m^2\phi^2/2+\lambda\phi^4/4!,~\lambda>0$ where $m^2$ is the renormalized mass.
%
%In this case, the set of critical points lie along the curve $m^2=0, \lambda>0$ because a local maximum at $\phi=0$ and two local minima at $\phi=\pm\sqrt{-6m^2/\lambda}$ for $m^2<0$ become degenerate at $m^2=0$.  
%
As we see below, the multiplicity of a criticality is equal to the number of fine tunings to reach it.

\begin{figure}
\begin{center}
\includegraphics[width=7cm]{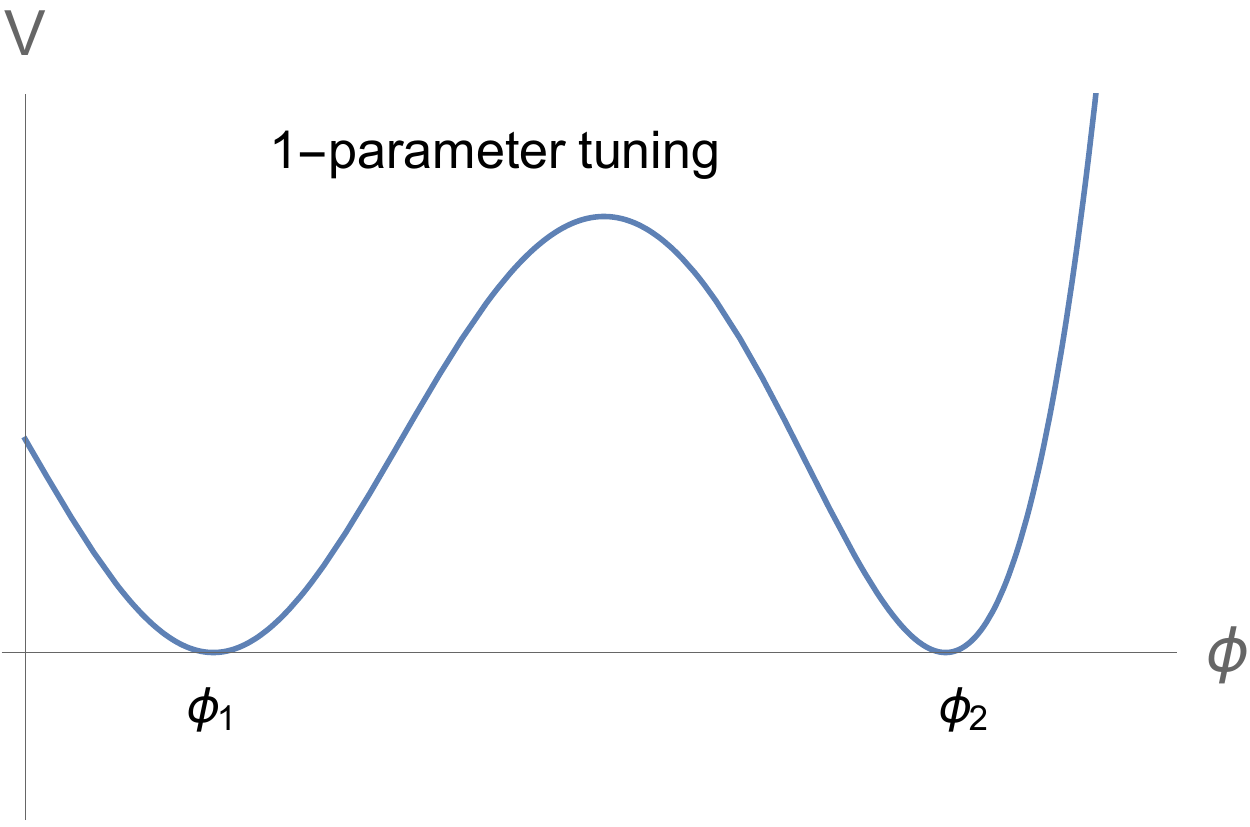}
\includegraphics[width=7cm]{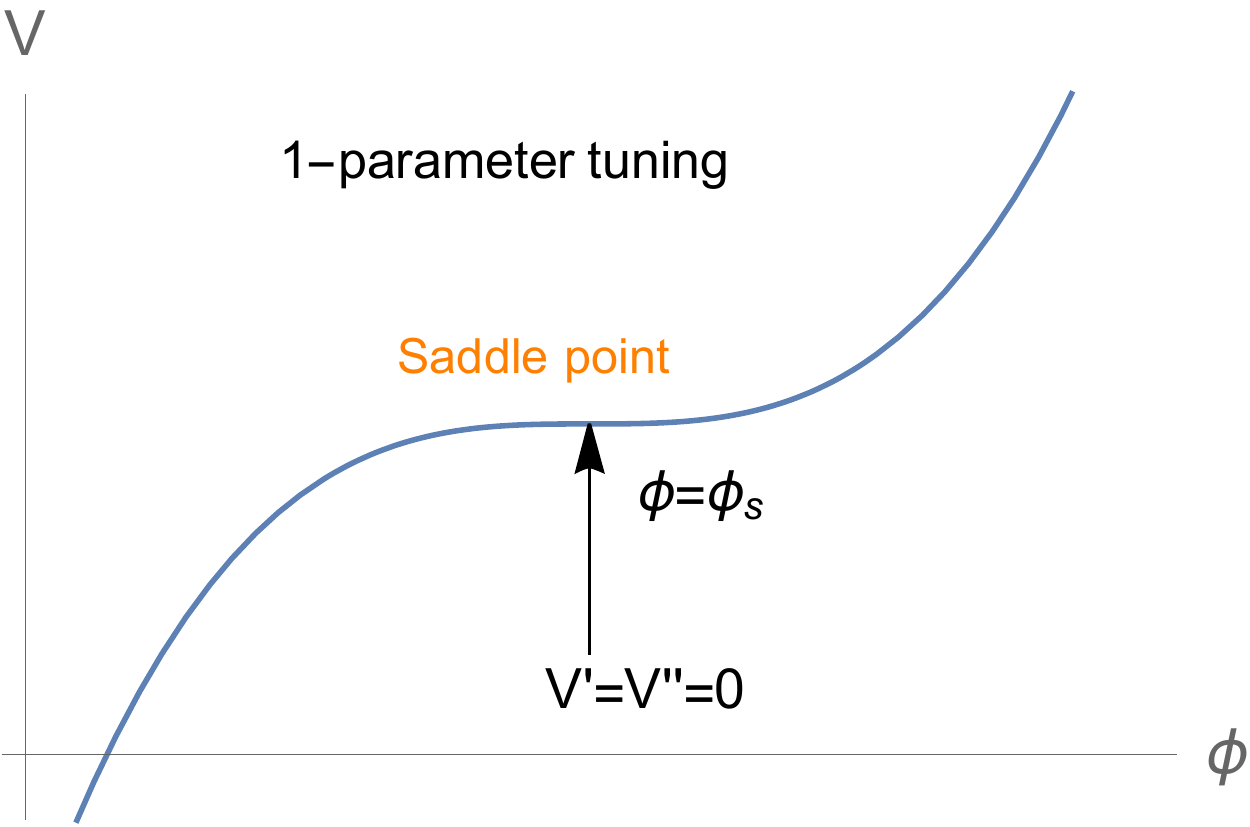}
\includegraphics[width=7cm]{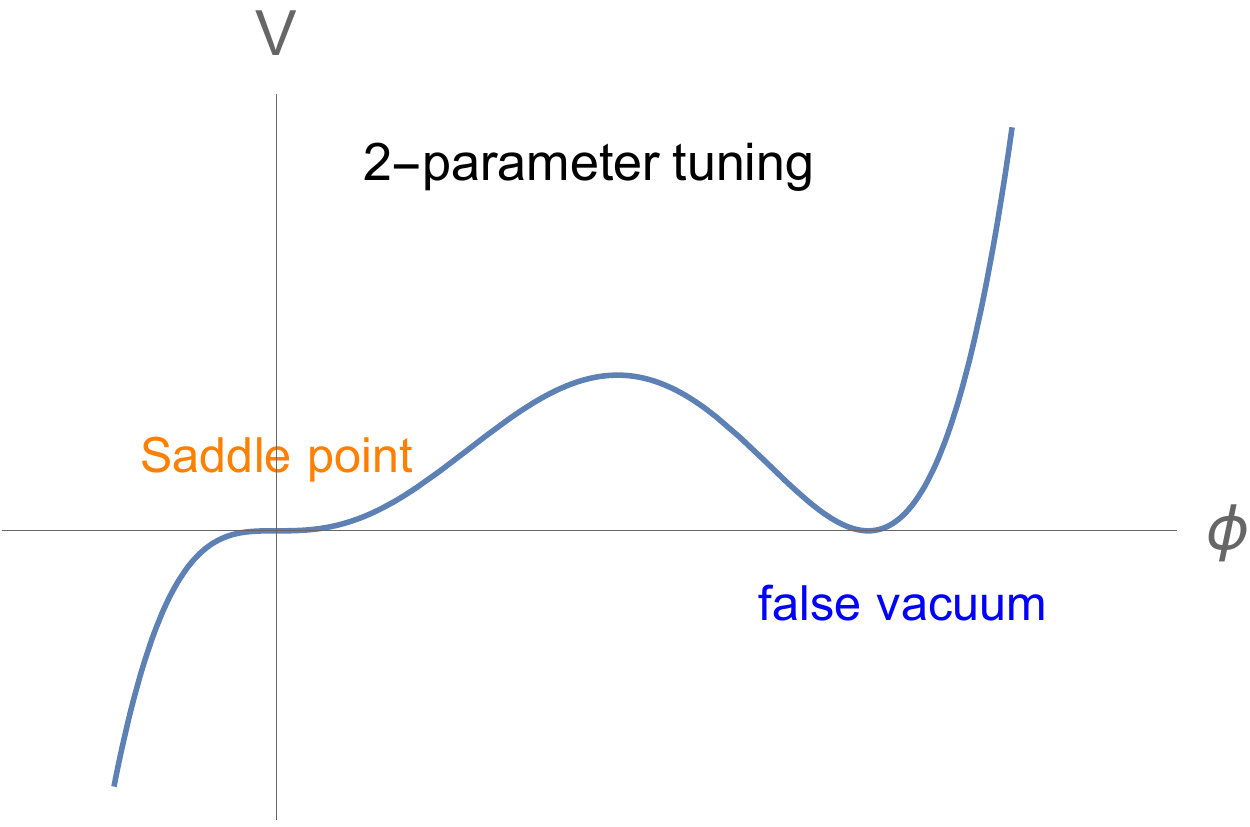}
\includegraphics[width=7cm]{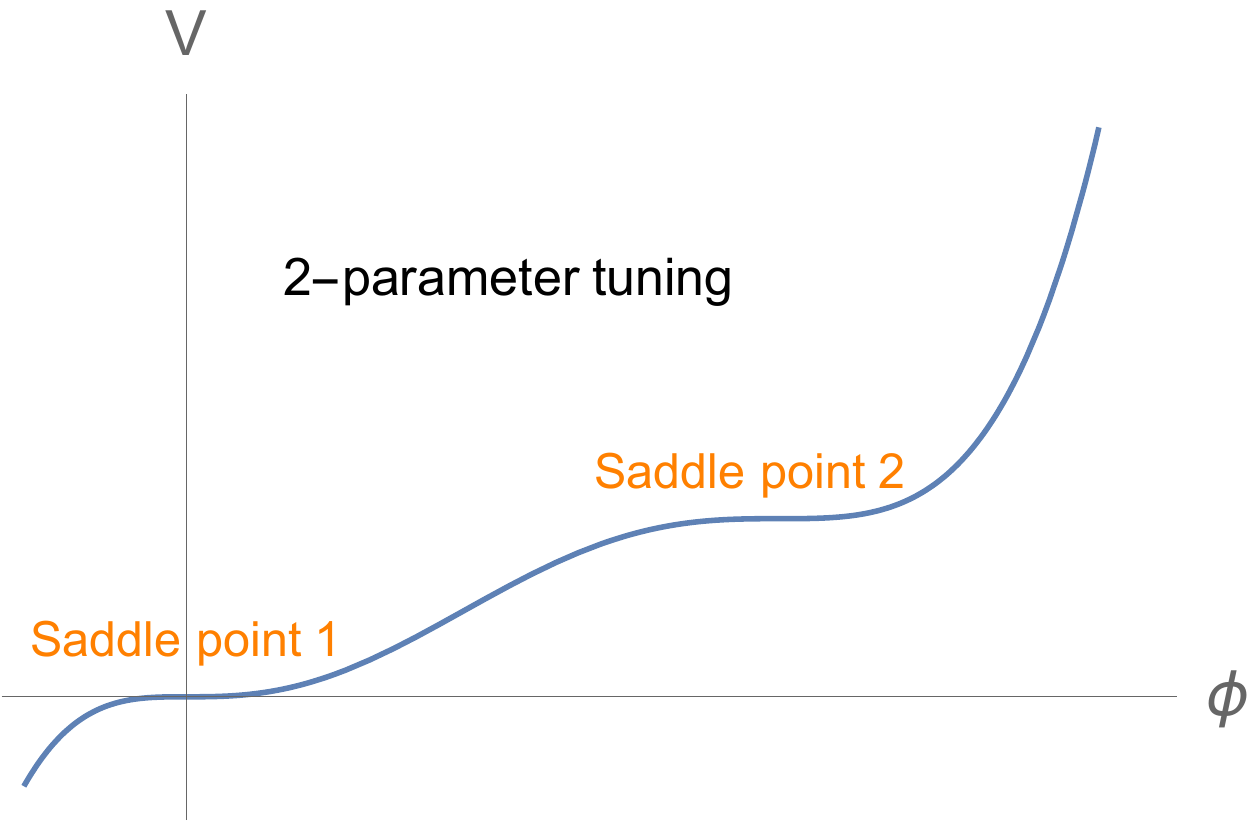}
\caption{Typical examples of the MPP. 
The upper panels correspond to 1-parameter tuning while the lower panels correspond to 2-parameter tuning.  
} 
\label{fig:MPP}
\end{center}
\end{figure}

The generic criticality is the degeneracy of two extrema,
which can be classified into two cases.
One is that two non-adjacent extrema are degenerate. 
(See the upper-left panel in Fig.~\ref{fig:MPP} for example.) 
The other is that two adjacent local minimum and local maximum degenerate to
form a saddle point.
(See the upper-right panel in Fig.~\ref{fig:MPP}.)
In either cases, the criticality is obtained when one equation is satisfied, that is, 
$V|_{\phi=\phi_1^{}}^{}=V|_{\phi=\phi_2^{}}^{}$ for the former or 
$V'|_{\phi=\phi_s^{}}^{}=V''|_{\phi=\phi_s^{}}^{}=0$ for the latter case.
Therefore the generic criticality is reached by one-parameter tuning.
We can also consider multiple criticalities that correspond to multi-parameter tunings.
The lower two panels in Fig.~\ref{fig:MPP} show examples of the doubly critical points. 
The lower-left panel  shows a double criticality where a local minimum (false vacuum) degenerates with a saddle point,
while in the lower-right panel there are two saddle points.
It is easy to check that they are obtained by tuning two parameters.

%
%
%}
%
In general, there are a number of multiple critical points in the (one-loop) effective potential.
The most important of these is the point with the highest criticality. This is because they are the most likely to occur according to the MPP. 
These critical points are called ``maximally critical points". 
As we will explain, one of them corresponds to the the classical conformality (CC) assumption~\cite{Meissner:2006zh,Foot:2007iy,Iso:2009ss,Iso:2009nw,Hur:2011sv,Iso:2012jn,Englert:2013gz,Hashimoto:2013hta,Holthausen:2013ota,Hashimoto:2014ela,Kubo:2014ova,Kubo:2015cna,Kawana:2015tka,Jung:2019dog,Jung:2021vap}, that leads to the Coleman-Weinberg (CW) potential.  
However, from the MPP point of view, all the maximally critical points are equally important, and the CW potential should have several variations corresponding to them.

The ultimate goal is to classify the maximally critical points of general theory.
If we restrict the problem to one-loop, we can write down 
the general form of the effective potential (see Eq.~(\ref{CW potential})). 
However the systematic search for the multi-critical points of such a  potential is still too complicated a problem.
%and less attractive from the phenomenological point of view.    
%the one-loop effective potential of a general real scalar field $\phi$.  
%In this paper, we study the multi-critical points of a general scalar field $\phi$ at one-loop level.  
%
%We can always write a general form of the one-loop effective potential (see Eq.~(\ref{CW potential})), but studying multi-critical points of it without any assumptions is too complicated and less attractive from the phenomenological point of view.    
%
Here, we will simply assume that the vacuum expectation values of all the fields except the field $\phi$ are zero. 
Furthermore, we assume that the other fields are either massless or so heavy that they do not contribute to the low energy effective potential of $\phi$. 
%only the $\phi$ sector has dimensionful parameters at tree level, which guarantees that all the background-dependent masses $M_i^{}(\phi)$ are proportional to $\phi$ except for $M_\phi^{}(\phi)$.    
%
We can imagine that the masslessness of the other fields appears as a consequence of 
the MPP as in the case of the CC.  
%and
Thanks to these assumptions, the one-loop effective potential can be expressed by four parameters (see Eq.~(\ref{CW potential 3})), and we can     
%From phenomenological point of view, we assume that one-loop part of the effective potential can realize the CW mechanism at around low-energy scale $M$.  
%is responsible for a symmetry breaking by the CW mechanism. 
%Within this assumption, 
study the maximally critical points in a model-independent manner.   
The study of doubly critical points corresponding to two-parameter tunings is also presented in \ref{doubly}.  
%Although our study does not cover whole the possibilities of multi-critical points of the (one-loop) effective potential, %it  still can be widely applicable to many models of the Coleman-Weinberg mechanism such as the models with the CC.    
%Of course, the precise phenomenological  predictions are model-dependent, but 
Although the effective potential we consider here is rather restricted,
the result can be used to analyze various models of the modification of the SM.
%we want to emphasize that our purpose here is not to give a comprehensive analysis of multi-critical points but
At any rate, the main purpose of this paper is to give general idea and philosophy of 
the MPP so that readers can use it when studying a concrete model of particle physics. 
%
%it still can be widely applicable to many models of the Coleman-Weinberg mechanism. 
%
%Our main results are summarized in Table.~\ref{tab:case}. 
%We note that existence of a saddle-point around the Planck scale, rather than a degenerate vacua, is another possible form of multicriticality. This fact is used in the critical Higgs inflation~\cite{Hamada:2014iga,Bezrukov:2014bra,Hamada:2014wna,Hamada:2014xka,Hamada:2017sga} explained below.
%Besides such interesting behavior of the Higgs potential, there are many mysteries and problems in particle physics and cosmology.   

%
In general, the effective potential can also have another vacuum or saddle point at high energy scale due to the RG running effects as in the case of the Higgs potential. 
%Though our main focus in this paper is to find multi-critical points of one-loop effective potential at low-energy scale, the effective potential can also have another vacuum or saddle point at high energy scale due to the RG running of quartic coupling as in the case of the Higgs potential. 
%
We will also briefly mention this possibility and its possible implications for cosmology. 
See Refs.~\cite{Hamada:2014wna,Bezrukov:2014bra,Ballesteros:2015noa,Ezquiaga:2017fvi,Lee:2020yaj,Hamada:2021jls,Cheong:2021vdb} and references therein for more details. 
%\blue{(Previous paper $\rightarrow$)
%In addition, the CMB fluctuations may also provide hints for further new physics since they are seeded at high energy scales during inflation. 
%
%Current observation is consistent with single-field inflation models, among which the Higgs inflation provides one of the best fits~\cite{Bezrukov:2007ep,Barvinsky:2008ia,DeSimone:2008ei,Allison:2013uaa}.  
%
%In particular, the critical Higgs inflation is inspired by the possible existence of the saddle point around the Planck scale in the MPP, which helps to flatten the Higgs potential and allows rather small value of the 
%non-minimal coupling $\xi={\cal O}(10)$ in $\xi |H|^2R$. (The not-so-large coupling is favorable from unitarity~\cite{Burgess:2009ea,Barbon:2009ya,Burgess:2010zq,Bezrukov:2010jz,Ema:2016dny}.)
%}

\

The organization of the paper is as follows. 
In Sec.~\ref{sec:potential}, we review one-loop effective potential.   
In Sec.~\ref{sec:criticality}, 
we classify the maximally critical points of the above mentioned effective potential. 
%We also give some discussions on the critical points with less multiplicity.
%we study the multi-critical points of one-loop effective potential by assuming that only the $\phi$ sector has dimensionful parameters at tree level.   
%
%In Sec.~\ref{inflation}, we discuss the critical Higgs inflation. 
%
%In Sec.~\ref{Prediction on inflationary observables}, we show the method and results for our numerical prediction for the inflationary observables.
%
Summary and discussion are given in Sec.~\ref{summary}. 
In \ref{doubly}, we discuss doubly critical points.

%___________________________________________
\section{One-loop effective potential}\label{sec:potential}
In this section, we discuss the one-loop effective potential of a general scalar field $\phi$. 
In a mass independent renormalization scheme, the renormalized one-loop effective potential is generally given by 
\aln{
V(\phi,\lambda_a^{},m^2;\mu)=\lambda_1^{}\phi+\frac{m^2}{2}\phi^2+\frac{\lambda_3^{}}{3!}\phi^3+\frac{\lambda_\phi^{}}{4!}\phi^4+\sum_i n_i^{}(-1)^{2s_i^{}}\frac{M_i^{}(\phi)^4}{64\pi^2}\ln \left(\frac{M_i^{}(\phi)^2}{\mu^2e^{C_i^{}}}\right)~, 
\label{CW potential}
}
where $s_i^{}$ and $n_i^{}$ represent the spin and the number of degrees of freedom (dof) of particle species $i$, respectively, $M_i^{}(\phi)$ is the background-dependent mass, and $\mu e^{C_i^{}/2}$ is the renormalization scale with $C_i^{}$ being the scheme dependent constant.\footnote{For instance, in the $\overline{\rm MS}$ scheme, $C_i^{}$ are $3/2$, $3/2$ and $5/6$ for scalars, fermions and gauge bosons, respectively.
}  
The renormalization-scale independence of the effective potential is expressed as
\aln{{\cal D}V(\phi;\mu)&:=\frac{d}{d\ln \mu}V(\phi;\mu)
\nn
&=\left(\mu\frac{\partial}{\partial \mu}+\sum_a \beta_{\lambda_a^{}}^{}\frac{\partial}{\partial \lambda_a^{}}-\gamma_m^{}m^2\frac{\partial}{\partial m^2}-\gamma_\phi^{}\phi\frac{\partial}{\partial\phi}
\right)V(\phi;\mu)=0~,
}
where $\lambda_a^{}$ and $\beta_{\lambda_a^{}}^{}$ denote a general coupling constant and its beta function, and $\gamma_\phi^{}$ is the wave function renormalization. 
The solution of this partial differential equation is \cite{Bando:1992wy}
\aln{
V(\phi,\lambda_a^{},m^2;\mu)=V(\overline{\phi},\overline{\lambda}_a^{},\overline{m}^2;\overline{\mu}=e^t\mu)~,
}
where $\overline{\lambda}_a^{}$, $\overline{m}^2$ and $\overline{\phi}$ are the running quantities;  
\aln{
&\frac{d\overline{\lambda}_a^{}}{dt}=\beta_{\lambda_a^{}}^{}(\{\overline{\lambda}_a^{}\})~,\quad \overline{\lambda}_a^{}|_{\overline{\mu}=\mu}^{}=\lambda_a^{}~,
\\
&\frac{d\overline{m}^{2}}{dt}=-\gamma_m^{}\overline{m}^{2}~,\quad \overline{m}^{2}|_{\overline{\mu}=\mu}^{}=m^2~, 
\\
&\frac{d\overline{\phi}}{dt}=-\gamma_\phi^{}\overline{\phi}~,\quad \overline{\phi}|_{\overline{\mu}=\mu}^{}=\phi~. 
}
In the following, we will not put bars explicitly and interpret that all the quantities obey the above RGEs % according to the above RGEs
as we vary $\mu$. 
%

%\noindent$\bullet$ {\bf Coleman-Weinberg mechanism}\\
It would be educational to revisit the Coleman-Weinberg (CW) mechanism here. 
(Note that this case corresponds to the criticality {\bf 234-15} in the next section. )
When a model has no dimensionfull parameters at tree level, the mass $M_i^{}(\phi)$ is typically given by $M_i^{}(\phi)=g_i^{}\phi$ where $g_i^{}$ denotes a coupling constant. 
Then, the one-loop effective potential becomes 
\aln{
V(\phi;\mu)&=\frac{\lambda_\phi^{}}{4!}\phi^4+\sum_i n_i^{}(-1)^{2s_i^{}}\frac{M_i^{}(\phi)^4}{64\pi^2}\ln \left(\frac{M_i^{}(\phi)^2}{\mu^2e^{C_i^{}}}\right)
\\
&=\frac{\lambda_\phi^{}+\delta \lambda_\phi^{}}{4!}\phi^4+\frac{\beta_{\lambda_\phi^{}}^{}}{2\cdot 4!}\phi^4\log\left(\frac{\phi^2}{\mu^2}\right)~, 
\label{potential with no masses}
} 
where $\beta_{\lambda_\phi^{}}^{}$ is the one-loop beta function of $\lambda_\phi^{}$ without the contributions from the wave function renormalization, and  
\aln{\delta \lambda_\phi^{}=\sum_i n_i^{}(-1)^{2s_i^{}}\frac{3g_i^4}{8\pi^2}\ln \left(g_i^{2}/e^{C_i^{}}\right)~.
}
%
%For now, let's assume $\beta_{\lambda_\phi^{}}^{}>0$.  
%
The easiest way to understand CW mechanism is to choose the renormalization scale as $\mu=M$ where $M$ is the scale at which $\lambda_\phi^{}+\delta \lambda_\phi^{}$ vanishes;
\aln{
V(\phi;\mu=M)=\frac{\beta_{\lambda_\phi^{}}^{}}{2\cdot 4!}\phi^4\log\left(\frac{\phi^2}{M^2}\right)~. 
\label{CW choice}
} 
This potential has a minimum at $\phi=Me^{-1/4}$, which induces a symmetry breaking.    
Note that if we impose the initial conditions of the RGEs at a high-energy scale $\Lambda$, $M$ is rougly given by 
\aln{M\sim \Lambda  \exp\left(-\lambda_\phi^{}/\beta_{\lambda_\phi^{}}^{}\right)\bigg|_{\mu=\Lambda}^{}~,
}
which is a well-known result of dimensional transmutation.  
%from high-energy point of view. 

As another choice of the renormalization scale, $\mu\sim \phi$ is also frequently used because it corresponds to the resummation of leading-log terms \cite{Bando:1992wy,Iso:2018aoa}. 
In this case, the potential Eq.~(\ref{potential with no masses}) becomes
\aln{V(\phi;\mu=\phi)=\frac{\lambda_{\phi}^{\rm eff}(\mu=\phi)}{4!}\phi^4~,
\label{mu-phi choice}
}   
where $\lambda_{\phi}^{\rm eff}:=\lambda_\phi^{}+\delta\lambda_\phi^{}$.  
%
%Thus, the information of one-loop corrections is contained in the running effective coupling $\lambda_{\rm eff}^{}(\phi)$. 
%
This choice is particularly useful when we study large-field behaviors of the potential because no large logarithmic terms appear by definition. 
As a consistency check, it would be meaningful to reproduce the CW mechanism in this case:   
By assuming that $\lambda_{\rm eff}^{}(\phi)$ becomes zero at $\phi=M$, $\lambda_{\phi}^{\rm eff}(\phi)$ can be  approximately expanded as 
\aln{
\lambda_\phi^{\rm eff}(\phi)\sim 0+\frac{\beta_{\lambda_\phi^{}}^{}}{2}\log\left(\frac{\phi^2}{M^2}\right) 
}
around $\phi=M$, and we can see that Eq.~(\ref{mu-phi choice}) is now identical to Eq.~(\ref{CW choice}). 
This also confirms the renormalization-scale independence of the CW mechanism.     
%
%Eq~.(\ref{mu-phi choice}) has a minimum at
%\aln{V'=0\quad \Leftrightarrow\quad 0=4\lambda_{\rm eff}^{}+\frac{d\lambda_{\rm eff}^{}}{d\ln\phi}\sim 4\lambda+\beta_\lambda^{}~. }
In the following, we will mainly use $\mu=M$ since we are interested in the effective potential at low-energy scales.  
%

%___________________________________________________
\section{Multi-critical points of one-loop effective potential}\label{sec:criticality}
In this section, we study the critical behaviors of the one-loop effective potential Eq.~(\ref{CW potential}). 
%
%In particular, we want to locate multi-critical points at which different extrema degenerate each other. 
%
%Let us first clarify the relevant parameters in the one-loop effective potential $V(\phi;\mu)$.  
%
In a general gauge theory, in addition to $\phi$, we have other scalar field(s) $S$, gauge field(s) $A_\mu^{}$, and fermion(s) $\psi$.
% 
%At renormalizable level, 
Their masses that depend on the back-ground of $\phi$ %of scalar field $S$, gauge field $A_\mu^{}$, and fermion $\psi$ 
are given by
\aln{
M_\phi^{}(\phi)^2=m^2+\lambda_3^{}\phi+\lambda\phi^2/2~,\ M_S^{}(\phi)^2=m_S^2+\Lambda_{\phi S}^{}\phi+\lambda_{\phi S}^{}\phi^2/2~,\ M_A^{}(\phi)=g\phi~,\ M_\psi^{}(\phi)=m_\psi^{}+y\phi~,
\label{phi dependent mass}
}
respectively. 
%
%
%If we expand $M_i^{}(\phi)^4$ in Eq.~(\ref{CW potential}) with respect to $\phi$ , in general
%we have
%\aln{V(\phi;\mu)=&\phi\left[\sum_i a_{i}^{(1)} n_i^{}(-1)^{2s_i^{}}\log\left(\frac{M_i^{}(\phi)^2}{\mu^2e^{C_i^{}}}\right)\right]
%
%+\frac{\phi^2}{2}\left[m^2+\sum_i a_{i}^{(2)}n_i^{}(-1)^{2s_i^{}}\log\left(\frac{M_i^{}(\phi)^2}{\mu^2e^{C_i^{}}}\right)\right]
%\nn
%+\frac{\phi^3}{3!}&\left[\lambda_3^{}+\sum_i a_{i}^{(3)} n_i^{}(-1)^{2s_i^{}}\log\left(\frac{M_i^{}(\phi)^2}{\mu^2e^{C_i^{}}}\right)
%\right]
%+\frac{\phi^4}{4!}\left[\lambda^{}+\sum_i a_{i}^{(4)} n_i^{}(-1)^{2s_i^{}}\log\left(\frac{M_i^{}(\phi)^2}{\mu^2e^{C_i^{}}}\right)
%\right]~,
%+\frac{\lambda+\delta \lambda}{4!}\phi^4+\frac{c}{2\cdot 4!}\phi^4\log\left(\frac{\phi^2}{\mu^2}\right)+\frac{M_\phi(\phi)^4}{64\pi^2}\log\left(\frac{M_\phi^{}(\phi)^2}{\mu^2e^{3/2}}\right)~. 
%\label{general effective potential}
%}
%where $a_{j}^{(i)}$ are (dimensionful) coefficients determined by the coupling constants of the model.  
%
Studying the critical behavior of Eq.~(\ref{CW potential}) in this most general form is beyond the scope of this paper.  
Instead, we simply assume that only the $\phi$ sector has dimensionful parameters at tree level.   
%
%This simplification is sufficient for actual model buildings of dimensional transmutation by CW mechanism.  
%For simplicity, we assume that only $\lambda_1^{}$, $m^2$ and $\lambda_3^{}$ are the dimensionful parameters of a model.  
%
Under this assumption, Eq.~(\ref{phi dependent mass}) are reduced to
the form  $M_i^{}(\phi)=g_i^{}\phi$ except for $M_\phi^{}(\phi)$, % where $g_i^{}$ denotes a general coupling constant.  
and Eq.~(\ref{CW potential}) can be written as 
\aln{
V(\phi;\mu)=\lambda_1^{}\phi+\frac{m^2}{2}\phi^2+\frac{\lambda_3^{}}{3!}\phi^3+\frac{\lambda_\phi^{\rm eff}}{4!}\phi^4+\frac{c}{2\cdot 4!}\phi^4\log\left(\frac{\phi^2}{\mu^2}\right)+\frac{M_\phi(\phi)^4}{64\pi^2}\log\left(\frac{M_\phi^{}(\phi)^2}{\mu^2e^{3/2}}\right)~, 
\label{CW potential 2}
}
where $c$ is a coefficient determined by the coupling constants of the model, and the last term denotes the one-loop contribution from $\phi$ itself. 
The basic assumption of MPP is that the super-renormalizable coupling constants are adjusted to one of the multi-critical points. 
Therefore, we regard $c$ as a constant. 
%
%Here, we have also included the linear term of $\phi$ in order to make our study as general as possible. %\footnote{}  
%
In the following, we will neglect the last term and discuss the multi-critical points by the first five terms. 
%but it does not change the qualitative behaviors  of criticality, %as long as $m^2$ and $\lambda_3^{}$ are moderate values. 
%Thus, %Because our purpose here is not to give quantitative study, so we will neglect this term in the following discussion.  
%
This simplification is justified in the following sense: $c$ is typically ${\cal O}(\lambda^2,g^4,y^4)$, which is small as long as the model is perturbative. 
Then, the MPP fixes the remaining parameters $(\lambda_1^{},m^2,\lambda_3^{}%,\lambda_\phi^{}
)$ at small values $\sim c$, and this guarantees the smallness of the last term in Eq.~(\ref{CW potential 2}) because $M_\phi(\phi)^4\sim c^2\phi^4$, which is suppressed by a factor $c$ compared to the first five terms around the critical points.  
%${\cal O}(c^2\sim \lambda^4,g^8,y^8)$ at around the critical points. 
%

\ 

%___________________________________________________
\subsection{Multi-critical points at low-energy scale} 
Now let us investigate the critical behavior of Eq.~(\ref{CW potential 2}) in the low-energy region. 
By ``{\it low-energy}" here, we mean $\phi\sim M$ where $M$ is the CW scale where the effective quartic coupling $\lambda_\phi^{\rm eff}$ becomes zero. 
As in the case of the previous section, a simple way to study the effective potential around $\phi\sim M$ is to choose $\mu=M$. 
Then, we have
\aln{
V(\phi;\mu=M)&=\lambda_1^{}\phi+\frac{m^2}{2}\phi^2+\frac{\lambda_3^{}}{3!}\phi^3%+\frac{\lambda+\delta \lambda}{4!}\phi^4 
+\frac{c}{2\cdot 4!}\phi^4\log\left(\frac{\phi^2}{M^2}\right)%+\frac{M_\phi(\phi)^4}{64\pi^2}\log\left(\frac{M_\phi^{}(\phi)^2}{\mu^2e^C}\right)
~, 
\label{CW potential 3}
%\\
%&=M^4\left[\frac{\tilde{m}^2}{2}\tilde{\phi}^2+\frac{\tilde{\lambda}_3^{}}{3!}\tilde{\phi}^3%+\frac{\lambda+\delta \lambda}{4!}\phi^4 
%+\frac{c}{2\cdot 4!}\tilde{\phi}^4\log\left(\tilde{\phi}^2\right)\right]~,\label{CW potential 4}
}
where all the coupling constants are evaluated at $\mu=M$. 
%and all the dimensionful quantities are normalized by $M$ in the second line.    
%In this choice of the renormalization scale, the variation of $\lambda$ is now converted to that of $M$.
%
Note that $c$ should be positive to ensure the stability of the potential.

Here we should mention the higher-loop corrections. 
The MPP predicts that the tree-level couplings $\lambda_i^{}$ become the same order as $c$.  
Then, the two-loop corrections are typically ${\cal O}(\lambda_i^{3}/(16\pi^2))={\cal O}(c^3/(16\pi^2))\ll c$. 
% which are  quite small. 
%
Therefore, the critical points discussed in this paper are self-consistent in that the contributions of higher-loop corrections are negligible. 
%
%Of course,
More complicated critical points can appear when the tree-level couplings are of the same order as the coefficients of the higher-loop corrections. 
% can appear when including higher-loop corrections.
The study of such multi-critical points of general effective potentials is the next theoretical challenge.    

In Eq.~(\ref{CW potential 3}), the effective potential $V$ has
five parameters $\lambda_1^{}$, $m^2$, $\lambda_3^{}$, $c$, $M$. 
However, an overall rescaling of $V$ or $\phi$ does not change the pattern of the 
degeneracy of extrema. 
Therefore, we can fix $c$ and $M$ to some arbitrary values, 
and then examine the criticality by changing the other three parameters, $\lambda_1^{}$, $m^2$, $\lambda_3^{}$. 
In other words, we consider criticality in the three dimensional parameter space. Therefore a triple criticality is the maximum for this effective potential.
%
%and also that it is sufficient to focus on $\lambda_{1}^{}>0$ or $\lambda_3^{}>0$ due to the $Z_2^{}$  symmetry; $\lambda_{1}^{}\rightarrow -\lambda_{1}^{}~,\ \lambda_{3}^{}\rightarrow -\lambda_{3}^{}~,\   \phi\rightarrow -\phi$. 
% 

Next we examine the extremum points of $V$.
The derivatives of Eq.~(\ref{CW potential 3}) are 
 \aln{\frac{\partial V}{\partial \phi}&=\lambda_1^{}+m^2\phi+\frac{\lambda_3^{}}{2}\phi^2+\frac{c}{3!2}\phi^3\left[\ln\left(\frac{\phi^2}{M^2%e^{1/2}
 }\right)+\frac{1}{2}\right]~,
 \label{first derivative}
\\
\frac{\partial^2 V}{\partial \phi^2}&=m^2+\lambda_3^{}\phi+\frac{c}{2!2}\phi^2\left[\ln\left(\frac{\phi^2}{M^2%e^{1/2}
}\right)+\frac{7}{6}\right]~, \label{second derivative}
\\
\frac{\partial^3 V}{\partial \phi^3}&=\lambda_3^{}+\frac{c}{2}\phi\left[\ln\left(\frac{\phi^2}{M^2%e^{1/2}
}\right)+\frac{13}{6}\right]~, \label{third derivative}
\\
\frac{\partial^4 V}{\partial \phi^4}&=\frac{c}{2}\left[\ln\left(\frac{\phi^2}{M^2%e^{1/2}
}\right)+\frac{25}{6}\right]~. 
\label{fourth derivative}
}
By integrating these derivatives step by step, it is easy to see that there are generically five extrema. \footnote{Strictly speaking, depending on the coupling constants, the effective potential may have only three or one extremum. However, since triple criticality cannot be obtained in such cases, it is sufficient here to consider the generic case with five extrema. 
}
These are named 1, 2, 3, 4, and 5 from left to right on the $\phi$ axis, respectively. 
Note that 1, 3, 5 are local minima, while 2, 4 are local maxima. %\footnote{\red{Depending on parameters, it is also possible that the one-loop effective potential has only three (one) extrema. 
%
%Such cases are all included in our study  }}
%
%Note that $1~(2)$ and $5~(4)$ are essentially the same extrema due to the $Z_2^{}$ symmetry mentioned above. 
%\begin{figure}[t!]
%\begin{center}
%\includegraphics[width=10cm]{vacuum.pdf}
%\caption{Multi-critical points of the effective potential Eq.~(\ref{CW potential 3}). 
%} 
%\label{fig:vacuum}
%\end{center}
%\end{figure}
%

In the parameter space, for each pair of the extrema, there is a hypersurface on which those two extrema are degenerate.
This corresponds to a criticality obtained by a one-parameter tuning, and we express such a criticality as $ab$, where $a$ and $b$ are the name of the extrema that degenerate. 
Note that the degeneracy between neighboring extrema   (e.g. $12,~23,~\cdots$) corresponds to a saddle point. 

In principle, one can think of multi-critical points as the intersection of such hypersurfaces.
For higher critical points, however, it is easier to classify them in a combinatorial way. 
In fact, there are two categories of double critical points. 
In one category, three extrema $a$, $b$, and $c$ are degenerate, which is denoted as $abc$. 
In the other category, two pairs $(a,b)$ and $(c,d)$ are degenerate respectively, and it is denoted as $ab-cd$. 
Similarly, there are two categories of triple critical points. 
In one category, four extrema $a$, $b$, $c$, and $d$ are degenerate, which is denoted as $abcd$. 
In the other category, one triplet $(a,b,c)$ and one pair $(d,e)$ are each degenerate, which is denoted as $abc-de$. 
Note that a triple criticality consisting of three pairs like $ab-cd-ef$ does not exist in our case, because we have only five extremal points.

%and we represent a double(triple)-degeneracy as $123$, $234$, $345$, etc ($1234$, $2345$, $1235$, etc).  
%As we explained in Introduction, each of those contours corresponds to type-A or type-B criticality.  
%
%Moreover, it is also possible that different  degeneracies are happening at the same time, and we represent such a case as $12-45$, $14-35$, $123-45$, etc.   
%
%In Fig.~\ref{fig:vacuum}, we show those contours schematically where the different colors correspond to the different combinations of degeneracy.   
%
%ãã?ã?¡ã\textcircled{\scriptsize 3}ä»¥å€ã?ããããçŸè±¡è«çã«äœ¿ããã?%The intersecting points of those contours correspond to multi-critical points and we will discuss them one by one. 
%
In the following, we will explicitly write down the form of the effective potential at multi-critical points. 
We represent the true vacuum (false vacuum) as $\phi=v_\phi^{}~(v_f^{})$, the saddle point as $\phi_S^{}$, and the mass of $\phi$ at $\phi=v_\phi^{}$ as $m_\phi^{}$.  

\

\noindent $\bullet$ {\bf Maximally critical points}\\
\begin{figure}[t!]
\begin{center}
\includegraphics[width=5cm]{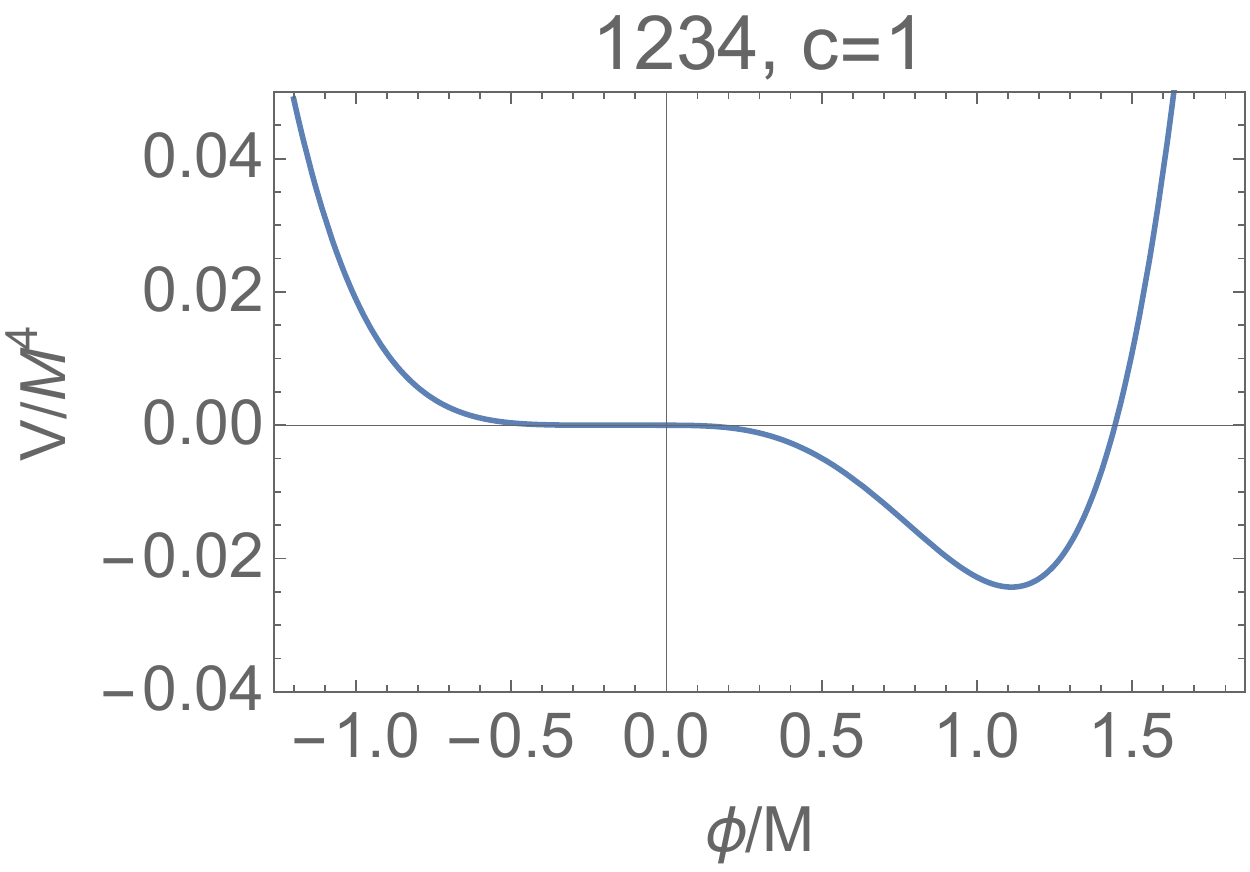}
\includegraphics[width=5cm]{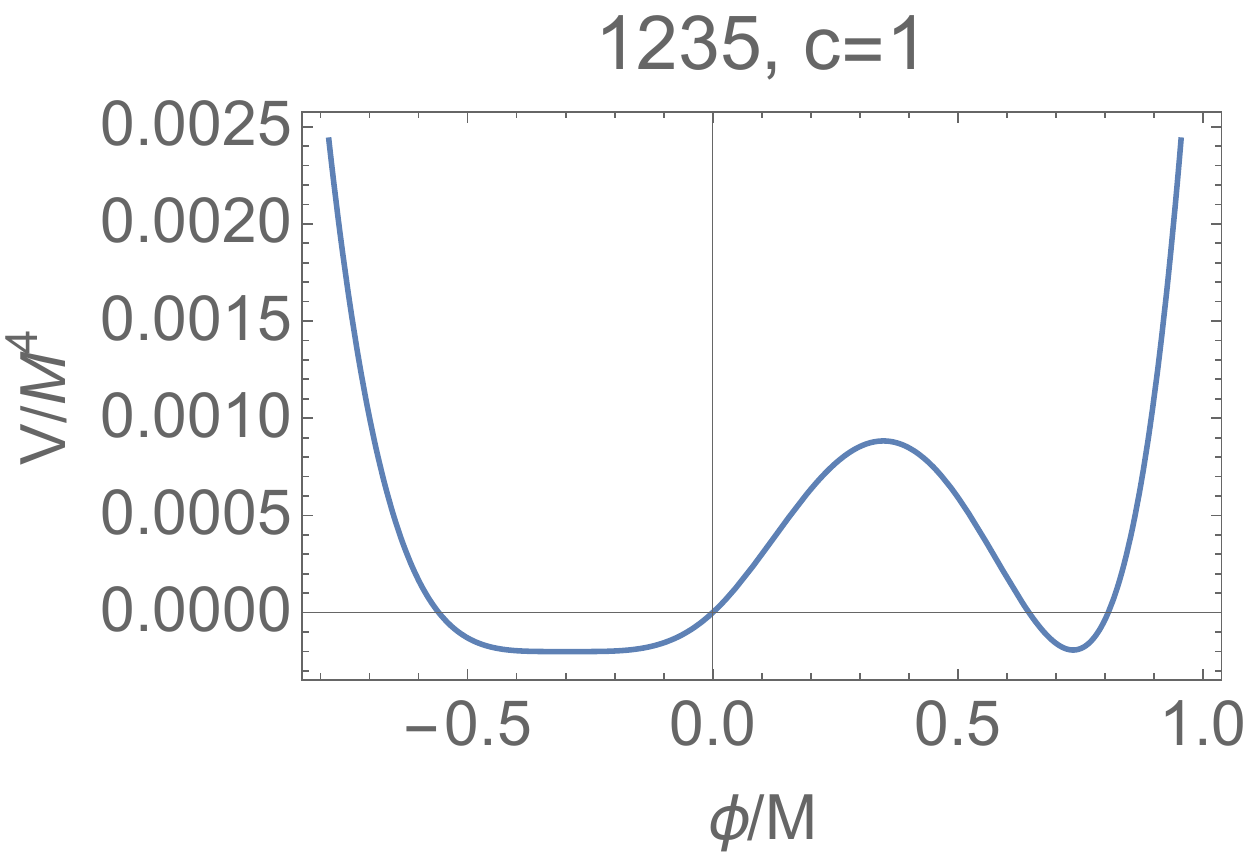}
\includegraphics[width=5cm]{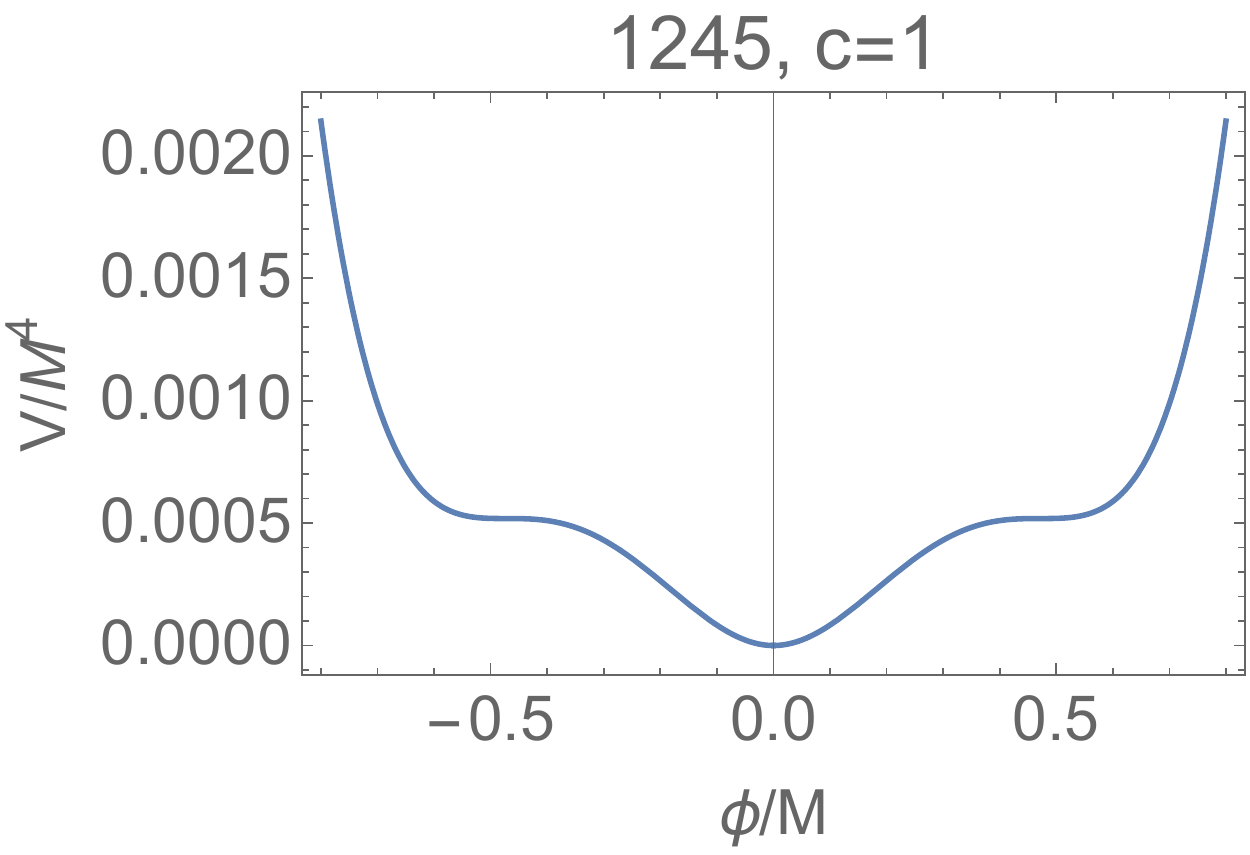}
\\
\includegraphics[width=5cm]{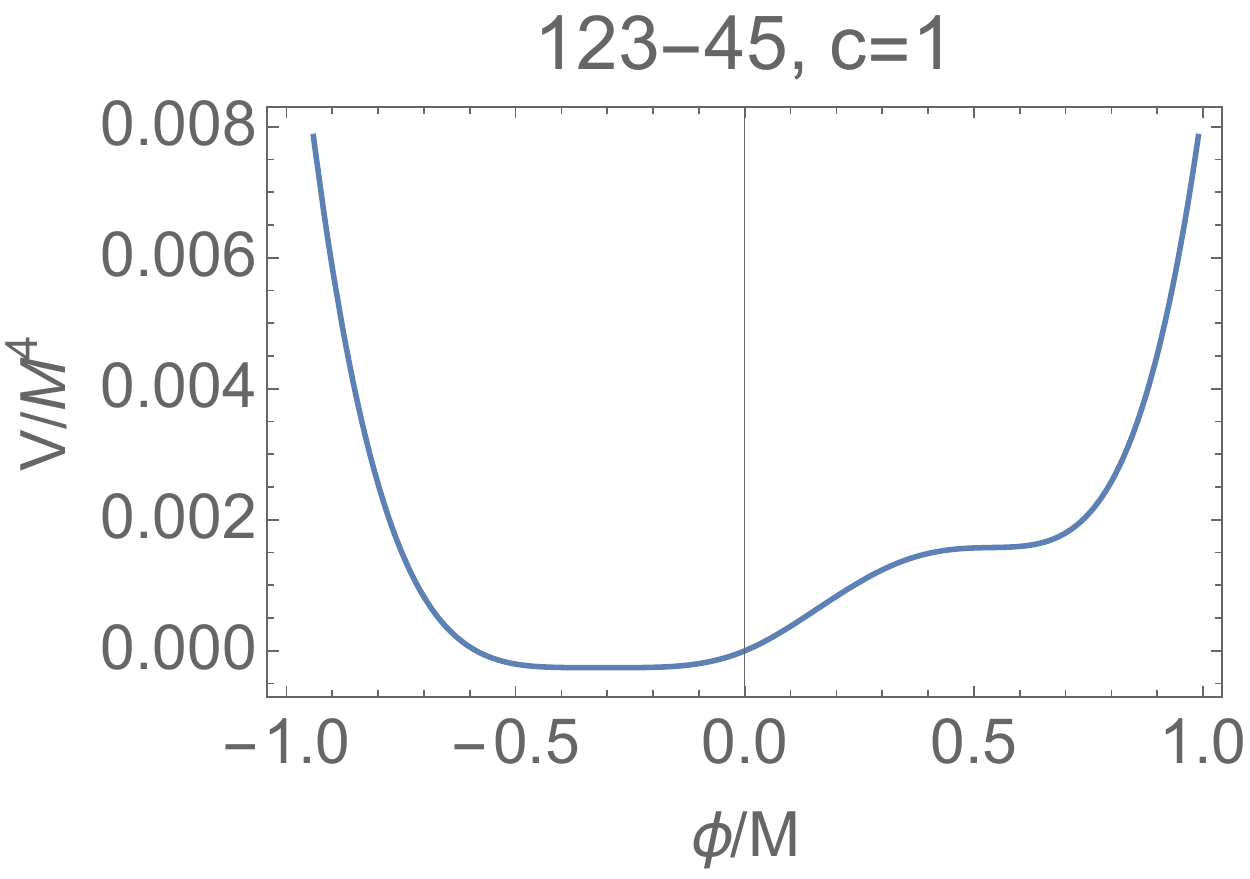}
\includegraphics[width=5cm]{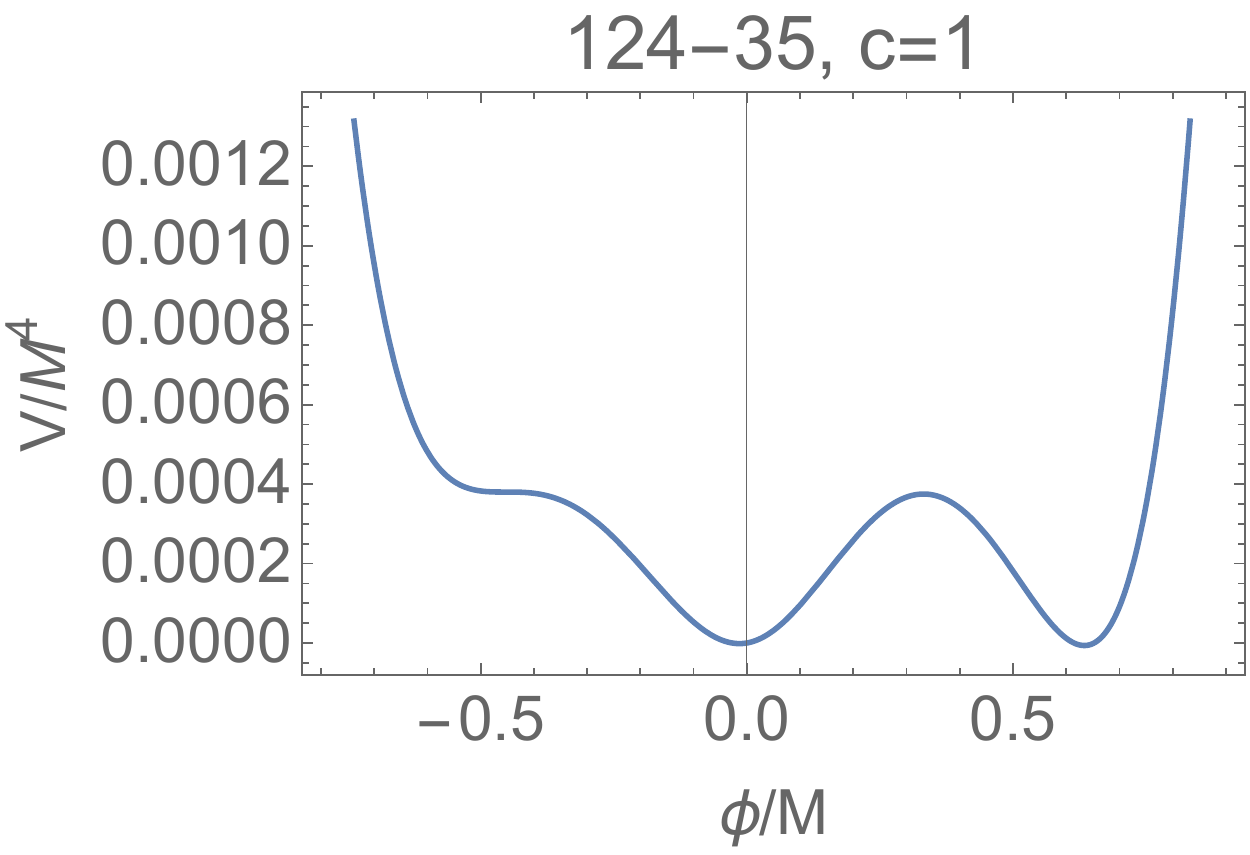}
\includegraphics[width=5cm]{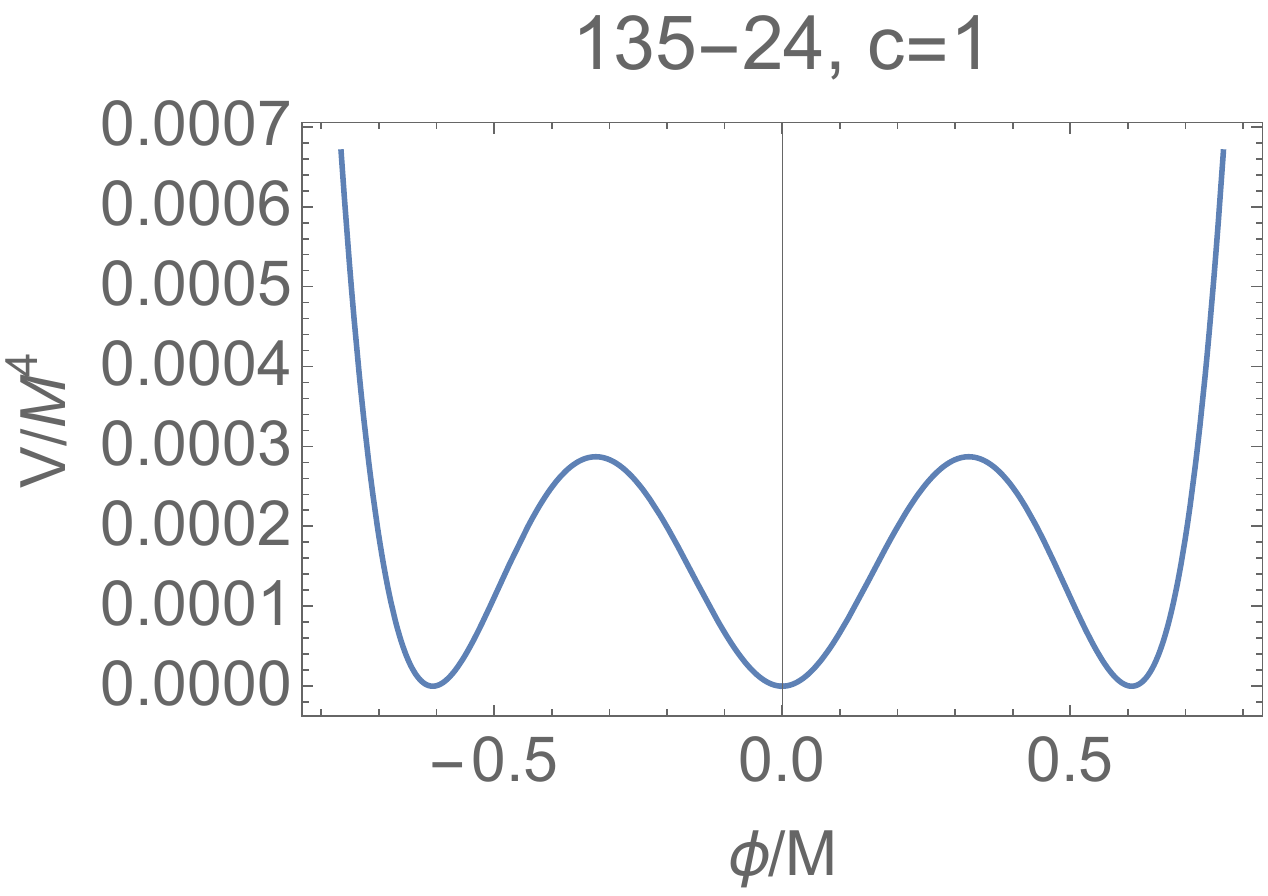}
\\
\includegraphics[width=5cm]{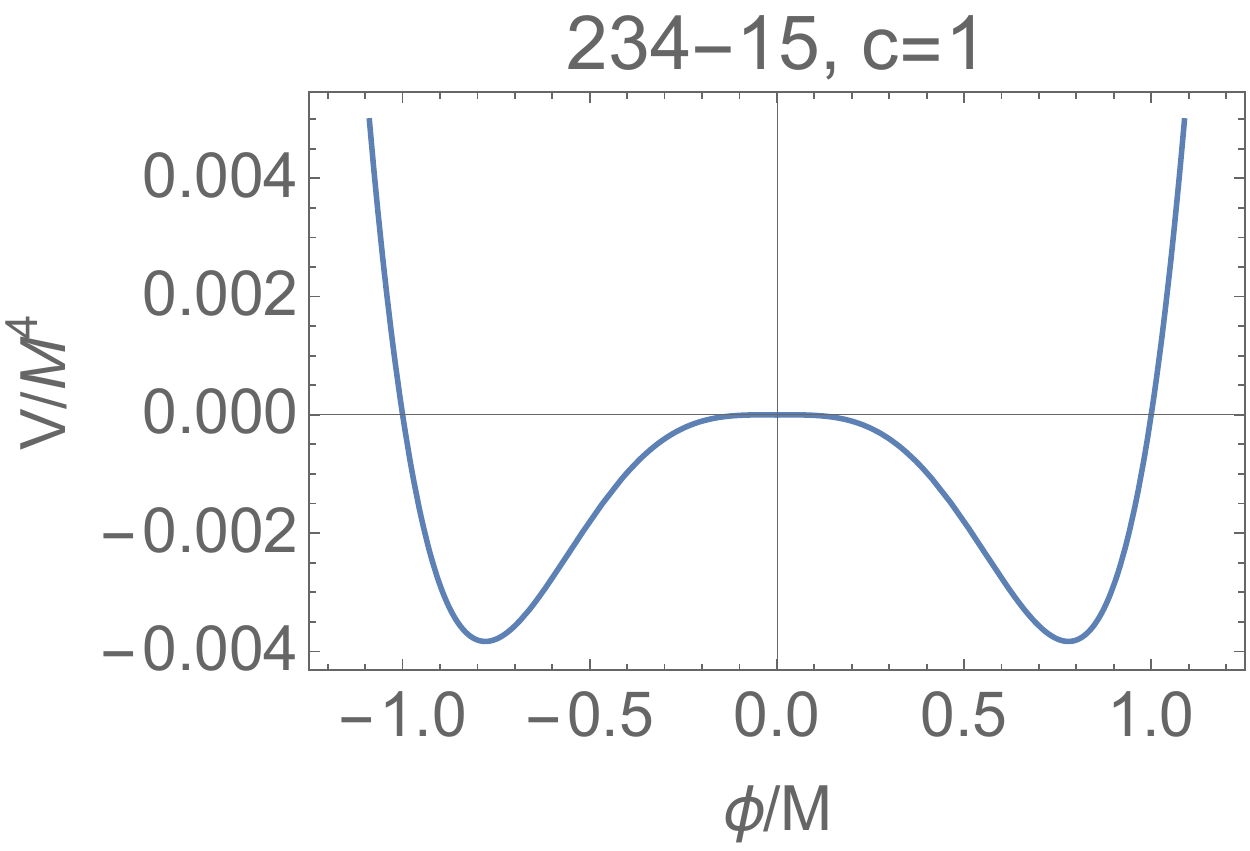}
\caption{
The effective potentials corresponding to the tricritical points. 
Here   
} 
\label{fig:tricritical}
\end{center}
\end{figure}
Let us start with classifying the maximally critical points, that is, triple criticalities. 
First, we point out that there is no need to distinguish between two criticalities that can be transferred by reversing the sign of $\phi$. 
For example, $ 123-45 $ and $ 543-21 $ are considered to be the same. 
Then the possible patterns of degeneracies are
\aln{& 1234~,\ 1235~,\ 1245~,%\ 1345~,\ 2345~, 
\nn
 123-45~,\ 124-35~,\ & 125-34~,\ 134-25~,\ 135-24~,\  234-15~.  
}
In the second line, however, not all of them are realized, which can be seen as follows.  
For example, let's take a look at $125-34$. 
Here, $V(a)$ denote the value of $V$ at the extremum $a$. 
Because $3$ is a neighbor of the local maximum $2$, we have $V(2)>V(3)$. 
Similarly, because $4$ is a neighbor of the local minimum $5$,  we have $V(4)>V(5)$. 
On the other hand from the degeneracy pattern we have $V(2)=V(5)$ and $V(3)=V(4)$,
which contradict with the above inequalities. 
Similarly, $134-25$ is not realized.

Thus, we find that there are seven triple critical points. As we have discussed, 
each of them needs a three-parameter tuning, and the effective potential 
is completely fixed up to overall rescalings of $V$ and $\phi$.
We give their explicit forms below. 

\
 
\noindent $\ast$ {\bf 1234}: This criticality is defined by 
\aln{V'=V''=V'''=V''''=0\quad {\rm at}\ \phi=\phi_S^{}~. 
}
By using Eq.~(\ref{fourth derivative}), $\phi_S^{}$ is given by
\aln{\phi_S^{}=\pm Me^{-25/12}~,
}
from which we can solve other conditions $V'|_{\phi=\phi_S^{}}^{}=V''|_{\phi=\phi_S^{}}^{}=V'''|_{\phi=\phi_S^{}}^{}=0$ as  
\aln{\lambda_1^{}=\pm \frac{c}{18}M^3e^{-25/4}~,\ m^2=-\frac{c}{4}M^2e^{-25/6}~,\ \lambda_3^{}=\pm cMe^{-25/12}~.
}
Then, the true vacuum $v_\phi^{}$ and $m_\phi^{}$ can be numerically solved as  
\aln{v_\phi^{}=\mp 1.1M~,\quad m_\phi^2=0.28cM^2~.
}
We show the resultant potential in the upper-left panel in Fig.~\ref{fig:tricritical} where $\lambda_1^{}<0$ is chosen.  

\

\noindent $\ast$ {\bf 1235}: In this case, the parameters can not be solved analytically, and numerical calculations are necessary.  
The degeneracy $123$ corresponds to $V'|_{\phi=\phi_S^{}}^{}=V''|_{\phi=\phi_S^{}}^{}=V'''|_{\phi=\phi_S^{}}^{}=0$, and $\lambda_1^{}$, $m^2$, and $\lambda_3^{}$ are given as functions of $\phi_S^{}$ by those conditions.   
Then, by putting them into $V(\phi)$, we can numerically  find the solution of $V(\phi=\phi_S^{})=V(\phi=v_\phi^{})$ by changing $\phi_S^{}$ and $v_\phi^{}$: 
\aln{
&\lambda_1^{}=\pm 2.3\times 10^{-3}cM^3~,\ m^2=0.016cM^2~,\ \lambda_3^{}=\mp 0.039cM~,
\\
&\phi_S^{}=\mp 0.30M~,\ v_\phi^{}=\pm 0.74M~,\ m_\phi^2=0.062cM^2~. 
}
The resultant potential is shown in the upper-middle panel in Fig.~\ref{fig:tricritical} where $\lambda_1^{}>0$ is chosen. 
Note that $\phi_S^{}$ can also become the true vacuum   and $m_\phi^{}=0$ in such a case.   

\

\noindent $\ast$ {\bf 1245}: In this case, the effective potential has a $Z_2^{}$ symmetry $V(-\phi)=V(\phi)$ because the criticality does not change via $1\leftrightarrow 5$ and $2\leftrightarrow 4$.  
As a result, we have $\lambda_1^{}=\lambda_3^{}=0$, and $m^2$ can be obtained by solving the saddle point conditions $V'|_{\phi=\phi_S^{}}^{}=V''|_{\phi=\phi_S^{}}^{}=0$ as functions of $\phi_S^{}$ and $m^2$. 
The results are   
\aln{
m^2=\frac{c}{12}M^2e^{-3/2},\quad \phi_S^{}=\pm Me^{-3/4}~,  
\label{sol1 1245}
}
from which $m_\phi^2$ is calculated as
\aln{m_\phi^2=\frac{\partial V^2}{\partial \phi^2}\bigg|_{\phi=0}^{}=\frac{c}{12}M^2e^{-3/2}~. 
\label{sol2 1245}
} 
The resultant potential is shown in the upper-right panel in Fig.~\ref{fig:tricritical}. 
This case is phenomenologically less attractive because we can not realize a symmetry breaking by $\phi$. 

\

\noindent $\ast$ {\bf 123-45}: The procedure of finding $123-45$ is similar to $1235$: We can first obtain $\lambda_1^{}$, $m^2$, and $\lambda_3^{}$ as functions of $v_\phi^{}$ by solving $V'|_{\phi=v_\phi^{}}^{}=V''|_{\phi=v_\phi^{}}^{}=V'''|_{\phi=v_\phi^{}}^{}=0$. 
Then, we can numerically find the solutions of $V'|_{\phi=\phi_S^{}}^{}=V''|_{\phi=\phi_S^{}}^{}$ as functions of $\phi_S^{}$ and $v_\phi^{}$:
\aln{
&\lambda_1^{}=\pm 2.9\times 10^{-3}cM^3~,\ m^2=0.020cM^2~,\ \lambda_3^{}=\mp 0.027cM~,
\\
&%\phi_S^{}=
v_\phi^{}=\mp 0.31M~,\ m_\phi^2=0~.
}
The resultant potential is shown in the middle-left panel in Fig.~\ref{fig:tricritical} where $\lambda_1^{}>0$ is chosen. 

\

\noindent $\ast$ {\bf 124-35}: This case also has to rely on numerical calculations.  
By the degeneracy $12$, we can first obtain $\lambda_1^{}$ and $m^2$ as functions of $\lambda_3^{}$ and $\phi_S^{}$. 
Then, by substituting them into $V(\phi)$, we can numerically find the critical point where $124$ and $35$ are simultaneously realized by changing $\lambda_3^{}$ and $\phi_S^{}$.  
The results are
\aln{
&\lambda_1^{}=\pm 2.3\times 10^{-4}cM^3~,\ m^2=0.017cM^2~,\ \lambda_3^{}=\mp 0.010cM~,\ \phi_S^{}=\mp 0.45M~. 
}
As for $v_\phi^{}$ and $m_\phi^{}$, we have two possibilities, $3$ or $5$:
\aln{(v_\phi^{},m_\phi^2)=\begin{cases} (\mp 0.014M,0.016cM^2) & {\rm for }\ 3
\\
 (\pm 0.63M,0.036cM^2) & {\rm for }\ 5
\end{cases}\quad.
} 
The resultant potential is shown in the center panel in Fig.~\ref{fig:tricritical} where $\lambda_1^{}>0$ is chosen.     

\

\noindent $\ast$ {\bf 135-24}: As well as $1245$, the effective potential has a $Z_2^{}$ symmetry in this case. 
Thus, we have $\lambda_1^{}=\lambda_3^{}=0$, and $m^2$ can be obtained by solving $V|_{\phi=v_\phi^{}}^{}=V'|_{\phi=v_\phi^{}}^{}=0$ as functions of $m^2$ and $v_\phi^{}$.  
The results are 
\aln{
m^2=\frac{c}{24}M^2e^{-1}~,\quad (v_\phi^{},m_\phi^2)=\begin{cases} (0,m^2)
\\
 (\pm Me^{-1/2},\frac{c}{12}M^2e^{-1})\end{cases}~,%\quad  
%\frac{c}{12}M^2e^{-1}~.
\label{parameters in 135}
}
We show the resultant potential in the middle-right panel in Fig.~\ref{fig:tricritical}.

\

\noindent $\ast$ {\bf 234-15}: This critical point is a trivial one, $\lambda_1^{}=m^2=\lambda_3^{}=0$, and the one-loop effective potential is simply given by the CW potential~\cite{Coleman:1973jx}:
\aln{
V(\phi)=\frac{c}{2\cdot 4!}\phi^4\log\left(\frac{\phi^2}{M^2}\right)~,
\label{potential 234-15}
}
which has a minimum at $v_\phi^{}=\pm Me^{-1/4}$. 
The mass of $\phi$ is
\aln{m_\phi^{2}=\frac{c}{6}M^2e^{-1/2}~.
\label{mphi 234-15}
}
The resultant potential is shown in the lower panel in Fig.~\ref{fig:tricritical}.  
If we regard $\phi$ as the Higgs field, this is nothing but the original CW mechanism~\cite{Coleman:1973jx}. 
However, as is well known, it does not fit the phenomenology. 

\

Note that the CW potential Eq.~(\ref{potential 234-15}) appears as a consequence of the MPP. It is sometimes said that the quadratic term $m^2 \phi^2$ is absent because of the CC, although it is not easy to define it precisely in the sense of quantum theory. 
On the other hand from the MPP point of view, it is explained as one of the
maximally critical points~\cite{Haruna:2019zeu,Hamada:2020wjh}. 
In other words, the other maximally critical points listed above are equally important, and they predict that the renormalized mass term should be the same order as the CW scale $M$. 
In this sense, all of the critical points can have a potential to solve the gauge hierarchy problem. 
%For example, if we consider a complex scalar field coupled to a $U(1)$ gauge field as in the
%original CW mechanism, the potential should be $U(1)$ symmetric. Then only the cases like
%1245, 135-24, 234-15 appear as the maximally critical points. (Imagine that the potential is $U(1)$
%symmetric instead of $Z_2$.) Therefore we have two new possibilities in addition to the original
%CW potential.
%
%
Here we make sure what the underling principle of the argument is.  
Actually, we can regard the MPP as the first principle or 
% itself is the principle behind such solution for hierarchy problem. 
%Or 
we can regard it as a consequence of a deeper principle such as 
%there may be a quantum theory behind the MPP based on Nielsen's 
the micro canonical quantization of Bennet-Froggatt-Nielsen and a maximum entropy principle based on quantum gravitational multiverses. 
In any case, if we start from the MPP and look at its consequences, we get a larger class of criticalities that automatically includes the classical conformity.

In Table.~\ref{tab:tricritical}, we summarize all the maximally critical points of the one-loop effective potential Eq.~(\ref{CW potential 3}).  
We also study non-maximally critical points in \ref{doubly}.
%Here, ``{\it Degeneracy}" means the type-A criticality,  i.e. $V|_{\phi=\phi_i^{}}^{}=V|_{\phi=\phi_j^{}}^{}~(i\neq j)$. 
%
%
\begin{table}[!t]
\begin{center}
\begin{tabular}{|c||ccc|}\hline
            &  $\mathbb{Z}_2 $  & $|v_\phi^{}|$    & $m_\phi^2$        
            \\\hline\hline
{\bf 1234} &  Broken     & $1.1M$        &  $0.28cM^2$  
 \\\hline
 {\bf 1235}  &  Broken   & $0.74M$ or $0.30M$        &  $0.062cM^2$ or $0$    
 \\\hline
{\bf 1245}  &  Exact  & $0$    &   $cM^2e^{-3/2}/12$           
 \\\hline
{\bf 123-45}   &  Broken & $0.31M$  &  $0$          
 \\\hline
{\bf 124-35} &  Broken  & $0.014M$ or $0.63M$  &  $0.016cM^2$ or $0.036cM^2$            
   \\\hline
{\bf 135-24} &  Exact  & $0$ or $Me^{-1/2}$    & $cM^2e^{-1}/24$ or $cM^2e^{-1}/12$         
    \\\hline
{\bf 234-15} &  Exact  & $Me^{-1/4}$    &  $cM^2e^{-1/2}/6$         
    \\\hline
\end{tabular}
\caption{
All the maximally critical points of the one-loop effective potential Eq.~(\ref{CW potential 3}).  
}
\label{tab:tricritical}
\end{center}
\end{table}

\

%___________________________________________________
%\subsection{A couple of examples}
%\red{
Finally, let us look at two specific examples in which the one-loop effective potential is of the form Eq.~(\ref{CW potential 2}).   
The first example is the (non-)Abelian Higgs model
\aln{
{\cal L}=-|D_\mu^{}\phi_i^{}|^2-V_{\rm tree}^{}(\phi)-\frac{1}{4g^2}\text{Tr}(F^{\mu\nu}F_{\mu\nu}^{})~,
}
where $D_\mu\phi_i^{}=\partial_\mu^{}\phi_i^{}-ig A_{a\mu}^{} {{T^a}_{i}}^j\phi_j^{}$, and
\aln{V_{\rm tree}^{}(\phi)=m^2(\phi^\dagger\phi)+\frac{\lambda_\phi^{}}{4}(\phi^\dagger\phi)^2~,
} is the tree-level potential.  
Note that $\lambda_1^{}$ and $\lambda_3^{}$ are forbidden due to the gauge invariance. 
In this case, the back-ground dependent mass of $A_\mu^a$ is given by
\aln{
{M_{ab}}^2(\phi)=g^2\phi^\dagger {T^a}^\dagger T^{b}\phi~,
}
and the one-loop effective potential for $\phi$ is given by a similar equation to Eq.~(\ref{CW potential 2}) with $\lambda_1^{}=\lambda_3^{}=0$.    
Thus, there are three maximally critical points; $1245$, $135-24$, and $234-15$.   

The second, more non-trivial example is a model consisting of two real 
scalars with $Z_2^{}$ symmetry in one scalar sector:
\aln{
{\cal L}=-\frac{1}{2}(\partial_\mu^{}\phi)^2-\frac{1}{2}(\partial_\mu^{}S)^2-V_{\rm tree}^{}(\phi,S)~,
}
where
\aln{
V_{\rm tree}^{}(\phi,S)=\lambda_1^{}\phi+\frac{m^2}{2}\phi^2+\frac{\lambda_3^{}}{3!}\phi^3+\frac{m_S^2}{2}S^2+\frac{\lambda_{\phi S}^{}}{4}\phi^2 S^2+\frac{\lambda_\phi^{}}{4!}\phi^4%+\frac{\lambda_{\phi S}^{}}{4}\phi^2S^2
+\frac{\lambda_S^{}}{4!}S^4~. 
}
Here we assume the $Z_2^{}$ symmetry for $S$, and 
the coefficient of $\phi S^2$ is set to zero by shifting $\phi$.  
In this case, the back-ground dependent mass of $S$ is given by
\aln{M_S^2(\phi)=m_S^2%+\Lambda_{\phi S}^{}\phi
+\frac{\lambda_{\phi S}^{}}{2}\phi^2~,
}
which clearly does not satisfy the assumption that leads to  Eq.~(\ref{CW potential 2}) when $m_S^2\neq 0$.   
%
%However, the vanishing of $m_S^2$ can be also regarded as another criticality in the $S$ direction; 
%
%First, as for $\Lambda_{\phi S}^{}$ we can always eliminate this by shifting $\phi$. 
% 
%When $m_S^2<0$, $S$ has (at least) two minima around the origin  and they become degenerate for $m_S^2\rightarrow 0$. 
%
However, for general $m_S^2$, the effective potential involves $\log(\phi^2+m_S^2)$ instead of $\log(\phi^2)$, and it changes significantly at $m_S^2=0$. In fact, when $m_S^2$ becomes negative it becomes complex near $\phi=0$.
Therefore, the vanishing of $m_S^2$ can also be regarded as another criticality.
In this sense, the maximally critical points discussed above can be regarded as special cases of the quadruple critical points in the parameter space of $\lambda_1^{}$, $m^2$, $\lambda_3^{}$, and $m_S^2$. 
Finding all the quadruple critical points is also an interesting problem, and one that we will consider in future research.    
%}

%___________________________________________________
\subsection{Multi-critical points at high-energy scale}
So far, we have considered the low-energy behavior of the (one-loop) effective potential
Eq.~(\ref{CW potential 2}). 
In general, the effective potential can have another extremum at higher energy scales.
This is indeed the case for the Higgs field, as is shown using the renormalization group equation 
(RGE) \cite{Hamada:2014wna,Hamada:2014xka,Hamada:2021jls}. 
Actually, the realization of a degeneracy between our vacuum ($v=246~$GeV) and such a high energy vacuum was the initial motivation for the MPP in Ref.~\cite{Froggatt:1995rt,Froggatt:2001pa,Nielsen:2012pu}, and it is worth noting that the top mass was predicted to be around $170~$GeV based on this idea. 
%
%To study high-energy behaviors of Eq.~(\ref{CW potential 2}), we should note that, 
In the following, we consider the Planck mass $M_{\rm Pl}^{}$ as such high energy scales. 
Note that the details of low-energy behavior of the effective potential Eq.~(\ref{CW potential 2}) are not important to determine the criticality at high-energy scale $\phi\sim M_{\rm Pl}^{}$. 
This is because the linear, quadratic, and cubic terms in the effective potential are negligible compared to the quartic term as long as the mass scale $M$ of the couplings $\lambda_1, m^2, \lambda_3^{}$ are small compared with $M_{\rm Pl}^{}$.
%
%In other words, only quartic term is important to study the vacuum structure at around $\phi\sim M_{\rm Pl}$.  
%
Contrary, the existence of another extremum at high-energy scale is determined by the behavior
of the effective couplings which is obtained by the RGEs. %including $\lambda_\phi^{}$. 
Thus, instead of giving model-dependent analyses, let us discuss general aspects of high-energy criticality and mention a couple of implications for cosmology. 
See Refs.~\cite{Hamada:2014wna,Bezrukov:2014bra,Ballesteros:2015noa,Ezquiaga:2017fvi,Lee:2020yaj,Hamada:2021jls,Cheong:2021vdb} and references therein for more details.   

When $\phi\gg M$, by choosing $\mu=\phi$, Eq.~(\ref{CW potential 2}) can be approximated by
\aln{
V(\phi)\sim 
\frac{\lambda_\phi^{\rm eff}(\phi)}{4!}\phi^4~,
\label{running quartic potential}
}
where $\lambda_\phi^{\rm eff}(\phi)$ is the effective quartic coupling which contains the loop corrections.    
Note that this expression is also valid even if higher loop corrections are taken into account.   
%
%To study the high-energy behaviors of the potential, the  choice $\mu=\phi$ gives a good approximation of exact one because it corresponds to the resummation of leading logarithmic terms \cite{}. 
%
Then the extrema of $V$ at high-energy scales are determined by
\aln{\frac{\partial V(\phi)}{\partial \phi}=\frac{1}{4!}\left(\frac{d\lambda_\phi^{\rm eff}}{d\ln \phi}+4\lambda_\phi^{\rm eff}\right)\phi^4=0~,
\label{dVdphi}
}
where $\beta_{\lambda_\phi^{\rm eff}}^{}:=d\lambda_\phi^{\rm eff}/d\ln\phi$ can be interpreted as an effective beta function of the  quartic coupling. 
The question is how such high-energy extrema degenerate themselves or with the low-energy extrema.
In the following, we see that there are essentially two kind of criticalities.
%
%Depending on the RG runnings of $\lambda_{\phi}^{\rm eff}$ and $\beta_{\lambda_{\phi}^{\rm eff}}^{}$, there are typically two possibilities.  

\

\begin{figure}
\begin{center}
\includegraphics[width=7cm]{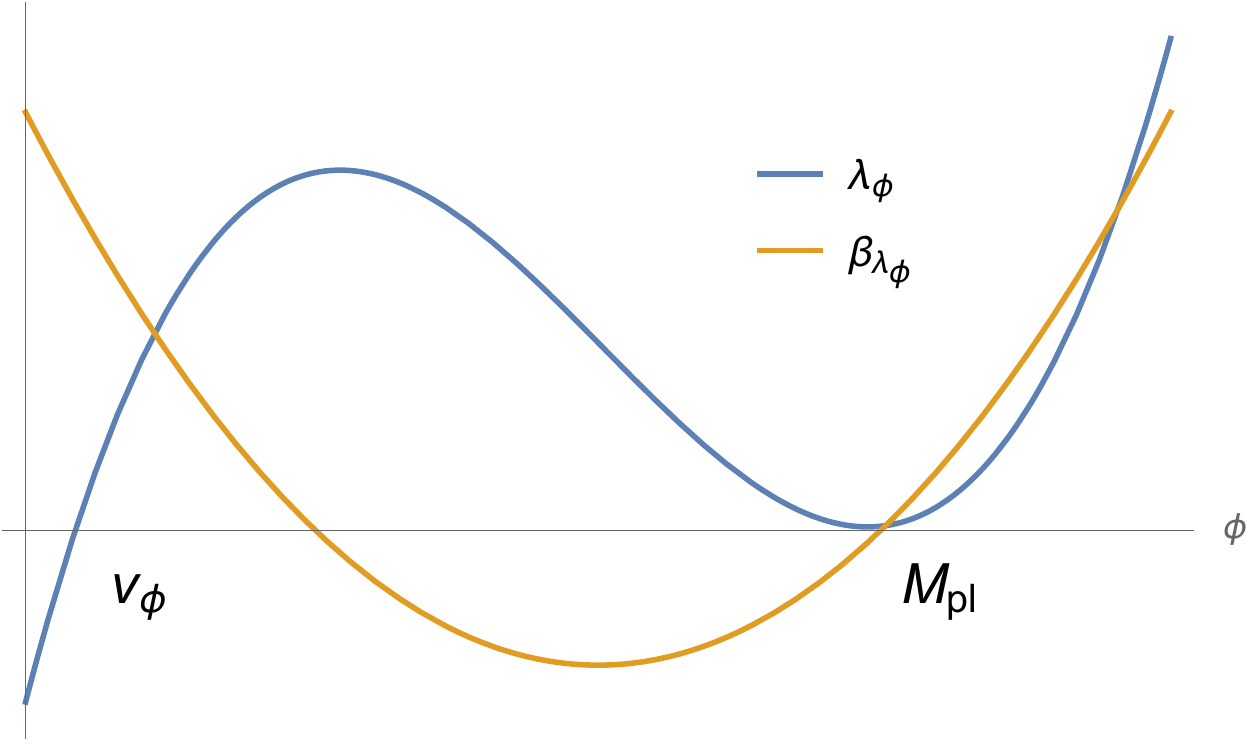}
\caption{Simultaneous realizations of the criticality $\lambda_\phi^{}\sim \beta_{\lambda_\phi^{}}^{}\sim 0$ at high-energy scale and the CW mechanism at low-energy scale. 
} 
\label{fig:typeA-high}
\end{center}
\end{figure}
\noindent$\ast$ {\bf Degeneracy with low-energy vacuum  
}\\
An extremum at high-energy scales can degenerate with a low-energy extremum, i.e. $V|_{\phi=v_\phi^{}}^{}=V|_{\phi=\tilde{v}_\phi}^{}$.
Here, $\phi=v_\phi^{}$ and $\phi=\tilde{v}_\phi$ are the low-energy and high-energy extrema, respectively.
% see once again Fig.~\ref{}. 
%
However, as long as $v_\phi^{}\sim M \ll \tilde{v}_\phi^{}$, $V|_{\phi=v_\phi^{}}$ is negligibly small
compared with the high-energy scales. 
Therefore the condition for the criticality is given by $V|_{\phi=\tilde{v}_\phi}^{}= V'|_{\phi=\tilde{v}_\phi}^{}= 0$.
From Eq.~(\ref{running quartic potential}) and Eq.~(\ref{dVdphi}), it is equivalent to 
\aln{\lambda_{\phi}^{\rm eff}= \beta_{\lambda_{\phi}^{\rm eff}}^{}= 0~,\quad \text{at $\phi=\tilde{v}_\phi^{}$}~. 
\label{mpp_1}
}
Note that this criticality is obtained by a one-parameter tuning of the coupling constants.
Suppose that $\lambda_{\phi}^{\rm eff}>0$ and $ \beta_{\lambda_{\phi}^{\rm eff}}^{}<0$ 
at some large value of $\phi~(<\tilde{v}_\phi^{})$. 
Then as $\phi$ is increased, $\lambda_{\phi}^{\rm eff}$ is expected
to go to zero at a certain value of $\phi$, $\phi=\tilde{v}_\phi^{}$.
Eq. (\ref{mpp_1}) requires that $ \beta_{\lambda_{\phi}^{\rm eff}}^{}$ goes to zero at the same time, which can be satisfied by adjusting one parameter.
It is also worth pointing out that if we want absolute stability of a low-energy vacuum,
$\lambda_{\phi}^{\rm eff}$ must be positive around $\phi\sim \tilde{v}_\phi^{} $.

%This conditions is typically satisfied when $\beta_{\lambda_{\rm eff}^{}}^{}<0$ for $\phi\lesssim \tilde{v}_\phi^{}$ because $\lambda_{\phi}^{\rm eff}$ must be positive in the intermediate region $v_\phi^{}\lesssim \phi\lesssim \tilde{v}_\phi^{}$ to guarantee the stability of the potential.  
%
It is a rather complicated problem to obtain a higher critical point by combining this criticality with
a low-energy criticality of the previous subsection.
As we have seen there, at low-energy scale $\phi\sim M$, $\beta_{\lambda_{\phi}^{\rm eff}}^{}$ needs to be positive in order to obtain a criticality.
%Namely, we have to realize
%\begin{numcases}{}
%\beta_{\lambda_{\phi}^{\rm eff}}^{}>0  & \text{for $\phi\sim v_\phi^{}$}
%\\
%\beta_{\lambda_{\phi}^{\rm eff}}^{}<0& \text{for $\phi\sim \tilde{v}_\phi^{}$}
%\end{numcases}
%
Thus, the beta function needs to flip its sign at some intermediate scale in order to archive simultaneous degeneracies among several low-energy extrema and a high-energy extremum. 
See Fig.~\ref{fig:typeA-high} for example. %a typical examplesrunning of $\lambda_{\phi}^{\rm eff}~(\beta_{\lambda_{\phi}^{\rm eff}}^{})$ by blue (orange).    
Such a criticality, sometimes called the ``{\it flatland scenario}"  \cite{Hashimoto:2013hta,Chun:2013soa,Ibe:2013rpa,Hashimoto:2014ela}, requires nontrivial RGEs and initial conditions of coupling constants for the sign of the beta function to flip.   
%  
%Thus, fermionic contributions to the beta function are  necessary to realize Eq.~(\ref{mpp_1}). 
%
%Moreover, 
%
Note that in the SM with the observed coupling constants, the Higgs potential can satisfy Eq.~(\ref{mpp_1}) near the Planck scale, while it is difficult to realize the low-energy criticality as in the CW mechanism at the weak scale. 

Let's investigate the high-energy extrema of $V$ in a more concrete manner.
First, we introduce an approximate expression of $\beta_{\lambda_{\phi}^{\rm eff}}$
around its zero point as
\aln{
\beta_{\lambda_{\phi}^{\rm eff}}^{}(\phi)\simeq b\ln \left(\phi/\mu_{\rm min}^{}\right)~,%\beta_s^{}+b\ln\left(\frac{\phi}{\phi_c^{}}\right)=b\ln\left(\frac{\phi e^{\beta_c^{}/b}}{\phi_c^{}}\right):=c\ln\left(\frac{\phi}{\mu_{\rm min}^{}}\right),\quad \mu_{\rm min}^{}:=\phi_c^{}e^{-\beta_c^{}/b}
\label{analytical beta}
} 
by regarding $d\beta_{\lambda_{\phi}^{\rm eff}}^{}/d\ln \phi:=b$ as a constant. 
In the following we assume $b>0$ for simplicity.
Obviously, $\beta_{\lambda_{\phi}^{\rm eff}}$ becomes zero at $\phi=\mu_{\rm min}^{}$.
From this we obtain
\aln{
\lambda_{\phi}^{\rm eff}(\phi)\simeq \lambda_{\rm min}^{}+\frac{b}{2}\left[\ln\left(\phi/\mu_{\rm min}^{}\right)\right]^2~,
\label{analytical lambda}
}
which takes the extremal value $\lambda_{\rm min}^{}$ at $\phi=\mu_{\rm min}^{}$.
Then from Eq.~(\ref{dVdphi}), the extrema of $V$, $\phi=\tilde{v}_\phi^{}$, are given by
\aln{\tilde{v}_\phi^{}=\mu_{\rm min}^{}\exp\left[\frac{1}{4}\left(-1\pm\sqrt{1-32\frac{\lambda_{\rm min}^{}}{b}}\right)\right]~. 
\label{vphitilde values}
}
From Eqs.~(\ref{running quartic potential})(\ref{analytical beta})(\ref{vphitilde values}), it is clear that both the number of extrema and the sign of the minimum value of $V$
depend on the value of $r:=\lambda_{\rm min}^{}/b~$:
When $r<1/32$, $V$ has one minimum and one maximum at high-energy scales.
In this region of $r$, the sign of the minimum value of $V$ is the same as that of $r$. 
When $r=1/32$, the minimum and maximum merge to form a saddle point, and they disappear when $r>1/32$. Obviously, the critical point we have considered above corresponds to $r=0$, i.e. $\lambda_{\rm min}^{}=0$.  

\

\noindent$\ast$ {\bf Saddle point criticality}\\
Another possible criticality involving high-energy extrema is that they themselves degenerate. 
Because there is generically one minimum and one maximum, the only possibility for criticality is for them to degenerate and form a saddle point.
 % $V'|_{\phi=\tilde{v}_\phi^{}}^{}=V''|_{\phi=\tilde{v}_\phi^{}}^{}=0$~. 
%
The vanishing condition of the first derivative is the same as Eq.~(\ref{dVdphi}), while the condition for the second derivative is
\aln{V''|_{\phi=\tilde{v}_\phi}^{}=0\quad \Leftrightarrow\quad
\left[12\lambda_\phi^{\rm eff}+7\beta_{\lambda_\phi^{\rm eff}}^{}+\frac{d\beta_{\lambda_\phi^{\rm eff}}^{}}{d\ln \phi}\right]\bigg|_{\phi=\tilde{v}_\phi^{}}^{}=0~.  
}
As long as $|d\beta_{\lambda_{\rm eff}^{}}^{}/d\ln \phi|\ll \lambda_\phi^{\rm eff},~|\beta_{\lambda_\phi^{\rm eff}}^{}|$ and $\lambda_\phi^{\rm eff}>0$, this condition also requires $\beta_{\lambda_{\rm eff}^{}}^{}<0$. 
Indeed within the approximation Eq.~(\ref{analytical lambda}), 
we have seen that we have a saddle point when $r=1/32$.
In other words, $V(\phi)$ has a saddle point when
\aln{\lambda_{\rm min}^{}=\frac{b}{32}%:=\lambda_c^{}
~,
}
and the location of the saddle point is $\tilde{v}_\phi^{}=\mu_{\rm min}^{}e^{-1/4}$.    

\

%________________________________________________________
\noindent $\bullet$ {\bf Implications for cosmology}\\
The existence of high-energy vacuum or saddle point has a lot of implications for cosmology. 
One of the well-known examples is the critical (Higgs) inflation \cite{Hamada:2014wna,Bezrukov:2014bra,Hamada:2014xka,Hamada:2017sga,Hamada:2021jls}. 
The inflation with non-minimal coupling $\xi \phi^2 R/M_{\rm Pl}^2$ typically needs large  $\xi\sim 10^{5}$  to explain the observed amplitude of the CMB fluctuations when $\lambda_\phi^{}={\cal O}(0.1)$. 
However, the MPP naturally favors smallness of $\lambda_\phi^{}\sim \beta_{\lambda_\phi^{}}^{}$ and this allows a successful inflation even when $\xi={\cal O}(10)$. 
% 
%The detailed inflation predictions are of course model-dependent. 
%We here give general analysis of saddle point inflation based on    

Moreover, the smallness of the quartic coupling is also important to discuss unitarity issue during the preheating stage after the inflation; 
As discussed in Refs.~\cite{Ema:2016dny,DeCross:2015uza,DeCross:2016fdz,DeCross:2016cbs,Sfakianakis:2018lzf,Ema:2021xhq}, the background dynamics of the inflaton shows a spike-like behavior around its  zero-crossings and can cause violent particle production of the Nambu-Goldstone modes of the inflaton or the longitudinal modes of the (weak) gauge bosons. 
In particular, the typical energy scale of those produced particles is ${\cal O}(\lambda_\phi^{1/2}M_{\rm Pl}^{})$, which is much larger than the cutoff scale $\Lambda:=M_{\rm Pl}^{}/\xi^{1/2}$ when  $\xi\sim 10^5$.~\footnote{This unitarity violation can be also understood from the point of view of strong coupling constants in the Einstein frame \cite{Hamada:2020kuy}.} 
Thus, the consistency (predictability) of the theory during the preheating stage is not clear. 
However, the situation changes quite a lot in the case of the critical (Higgs) inflation because the energy scale of produced particles is expected to be largely suppressed thanks to the smallness of $\lambda_\phi^{}$. % and it can become smaller than the cutoff scale $\Lambda$.  
%
%Therefore, even if it is consistent during inflation, the preheating dynamics still indicates the necessity of extending this model to a UV model which preserves unitarity.
The resultant cosmological predictions such as the reheating temperature need more detailed study of particle production and we want to discuss this possibility in some future investigation.

Another interesting application of high-energy criticality is to generate a seed of %large primordial density fluctuations for the seed of 
primordial black holes (PBHs) %at small scales  
\cite{Ezquiaga:2017fvi}.  
When overdense regions in the Universe have sufficiently  large density fluctuations $\delta \rho/\rho\gtrsim 0.4$, they can collapse into PBHs by overcoming the pressure of radiations. 
However, the observed fluctuations at large scales by the CMB is too small $\delta \rho/\rho\sim 10^{-5}$, and it is necessary to generate large fluctuations at small scales. 
Here, we can use the saddle point of the inflaton potential predicted by the MPP to obtain such large fluctuations.  
Naively, this can be seen by looking at the definition of the power spectrum of curvature perturbations in the slow-roll approximation
\aln{
{\cal P}_{R}^{}=\frac{V(\phi)^3}{24\pi^2 M_{\rm Pl}^6 V'(\phi)^2}~, 
\label{power spectrum}
}
%in the slow-roll approximation and 
which is proportional to $V'(\phi)^{-2}$. 
%
%where $U(\phi)$ is the inflaton potential in the Einstein frame. 
%  
However, detailed studies \cite{Germani:2017bcs,Drees:2019xpp,Liu:2020oqe} have shown that this is too naive because (i) we can not rely on the usual slow-roll approximations at around the saddle point and %the inflaton is passing the saddle point quickly due to the steep potential right before the saddle point and 
(ii) the slow-roll conditions are violated right before the saddle point. 
%during such an overshooting stage.  
%The existence of $U'(\phi)$ in the denominator in Eq.~(\ref{power spectrum}) indicates that the curvature perturbations can be largely enhanced when the inflaton potential $U(\phi)$ has a (near) saddle point at small scales. % while it must be $\sim 2.1\times 10^{-9}$ at large (CMB) scale. 
%
Although the predictions of the large density fluctuations in the critical (Higgs) inflation model are still in questioned, utilizing saddle point as a seed of large density fluctuations is an interesting possibility and naturally favored by the MPP. 

%%\begin{figure}
%\begin{center}
%\includegraphics[width=6cm]{PBH_1.pdf}
%\includegraphics[width=6cm]{PBH_2.pdf}
%\caption{The power spectrums of the curvature perturbation as functions of the e-folding number $N$ in the critical (Higgs) inflation.  
%} 
%\label{fig:PBH}
%\end{center}
%\end{figure}
%
%As examples, we show the power spectrums in the critical inflation case in Fig.~\ref{fig:PBH}, where we have used the approximate expression Eq.~(\ref{analytical lambda}) with
%\aln{\lambda_{\rm min}^{}=\lambda_c^{}(1+\delta)~,\quad c_s^{}=\frac{\tilde{v}_\phi^{}}{M_{\rm Pl}^{}/\sqrt{\xi}}~.
%}
%Here, $\delta$ denotes the deviation from the exact saddle-point criticality, and $c_s^{}$ is the ratio of the position of saddle point $\tilde{v}_\phi^{}$ to the cutoff $\Lambda=M_{\rm Pl}^{}/\sqrt{\xi}$.   
%
%Note that $\delta=0$ corresponds to the exact realization of a saddle point. 
%
%The horizontal axis is the e-folding number and the dashed orange lines correspond to the observed CMB amplitude $\sim 2.1\times 10^{-9}$. 
%
%One can see that the power spectrums have large half-dome peaks at small scales while they are consistent with the CMB amplitude at around $N=50-60$.  
%
%See Refs.~\cite{Ezquiaga:2017fvi,Kannike:2017bxn,Germani:2017bcs,Drees:2019xpp} for more detailed studies such as the mass of PBHs or their abundance.  

%__________________________________________________
\section{Summary and discussion}\label{summary}
In this paper, we have studied the multi-critical points of the one-loop effective potential. 
In order to simplify our discussion, we have assumed that the other fields except $\phi$ are either massless or too heavy to contribute to the low energy effective potential of $\phi$. 
Then, we have located all the maximally critical points in the parameter space of $\lambda_1^{}$, $m^2$, and $\lambda_3^{}$.  
One of these critical points is just the conventional CW potential, but the MPP also predicts other multi-critical points that allow for a symmetry breaking, and they should be equally treated in this framework.  
%
%Then, we have also discussed the critical points determined by 2-parameter tuning. 
%
%In this sense, the MPP admits a variety of generalizations of the CW mechanism.   
%Thanks to this assumption, the effective potential was parametrised by four effective couplings, and we were able to discuss the criticality model-independently. 
%
%Even within this assumption, our analysis can be applicable to many phenomenologically well studied models  
%From phenomenological points of view, this analysis seems to be sufficient since typical models of the CW mechanism 
%such as Ref.~\cite{Meissner:2006zh,Foot:2007iy,Iso:2009ss,Iso:2009nw,Hur:2011sv,Iso:2012jn,Englert:2013gz,Hashimoto:2013hta,Hashimoto:2014ela,Kawana:2015tka,Jung:2019dog,Jung:2021vap} where the classical conformality is usually assumed from the beginning.  
%are all belonging to this category, and our results can be applicable to such models. 
%
%
%In general, more complicated multi-critical points would appear when the contributions of massive fields and higher-loop corrections are taken into account.   
%
%The study of such multi-critical points of general effective potentials is a next theoretical subject.   

Phenomenological applications of these critical points are also interesting. 
They can have a lot of implications for cosmology since they can predict a (strong) first-order phase transition at the early universe. 
We want to report their consequences such as Gravitational Wave, electroweak baryogenesis, formation of primordial black holes etc. in the near future~\cite{kawana}.

%%%%%%%%%%%%%%%%%%%% ACKNOWLEDGMENTS %%%%%%%%%%%%%%%%%%%%
\section*{Acknowledgements} 
We would like to thank Yuta Hamada, Kinya Oda, and Kei Yagyu for useful discussions.   
The work of KK is supported by Grant Korea NRF-2019R1C1CC1010050, 2019R1A6A1A10073437.   
%

%%%%%%%%%%%%%%%%%%%%%%%%%%%Appendix%%%%%%%%%%%%%%%%%%%%%%%%%%%%%%%%%%%%%%%%
\appendix 
\def\thesection{Appendix \Alph{section}}
%_______________________________________________________________
%_______________________________________________________
%_______________________________________________________
%__________________________________________

%%%%%%%%%%%%%%%%%%%%%Non-maximally critical points%%%%%%%%%%%%%%%%%%%%%%%%%%%%%%%%%%%
\section{Doubly critical points}\label{doubly}
The critical points (lines) which correspond to 2-parameter tunings are less attractive from the point of view of the MPP. % because one of the coupling constants remains as a free parameter. 
% in the one-loop effective potential  Eq.~(\ref{CW potential 3}). 
%
Nevertheless, it would be meaningful to study them because  a lot of extended models are (automatically) categorized into this case. 
%
%For example, models with the CC typically belong to this category because linear and cubic terms of the effective potential are absent due to $M_i^{}(\phi)\propto \phi$.    
%
By considering the redefinition $\phi\rightarrow-\phi$, the possible degeneracies are given by
\aln{&123~,\ 124~,\ 125~,\ 134~,\ 135~,\ 234~,\ 
\\
 12-34~,\ 12-35~,\ 12-45&~,\ 13-24~,\ 13-25~,\ 14-23~,\ 14-25~,\ 15-23,\ 15-24~. 
} 
As in the case of the maximally critical points, it is easy to find that $14-23$ and $14-25$ are not realized by simply comparing the values of the extrema.   
\begin{figure}[t!]
\begin{center}
\includegraphics[width=5cm]{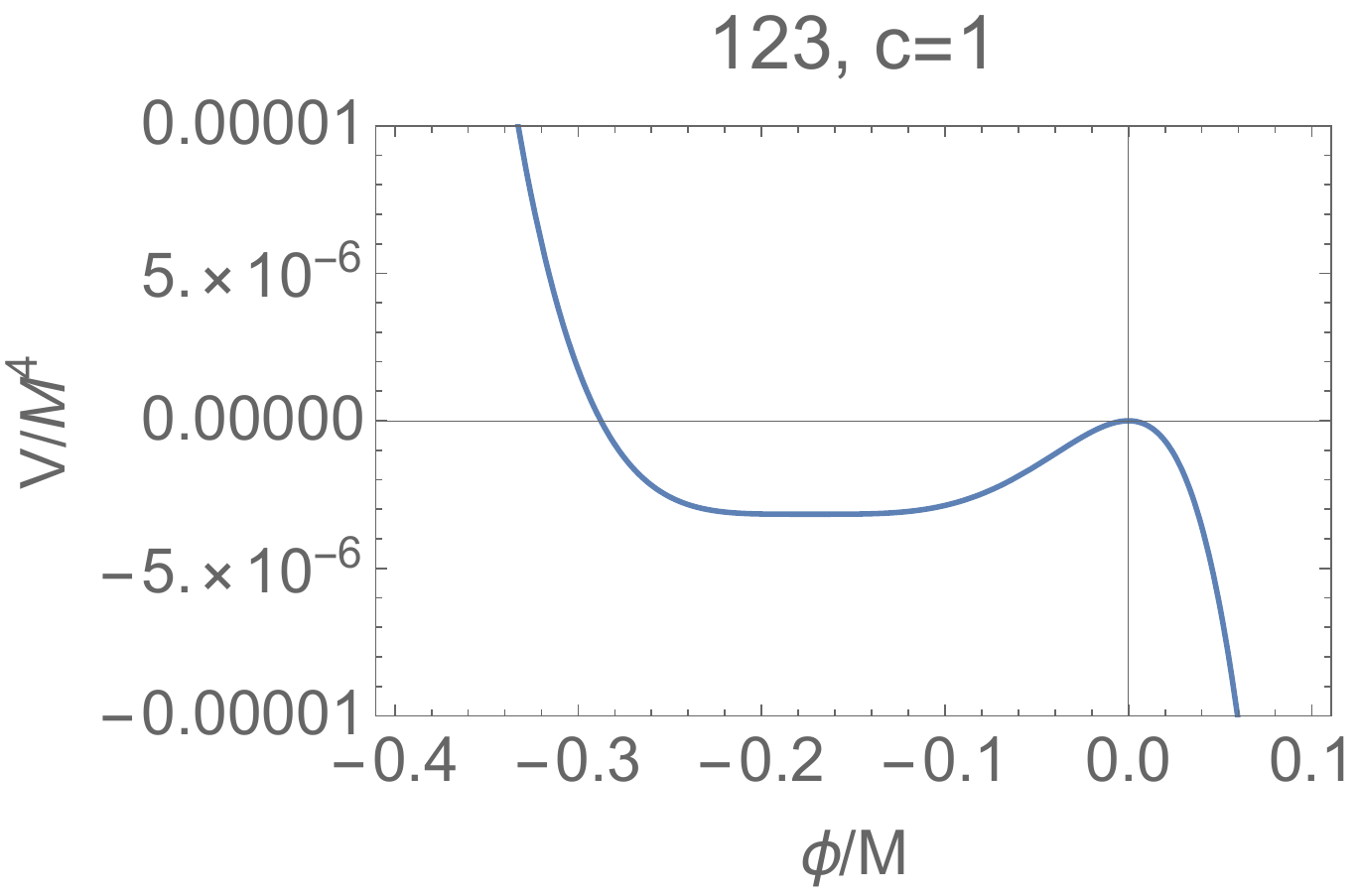}
\includegraphics[width=5cm]{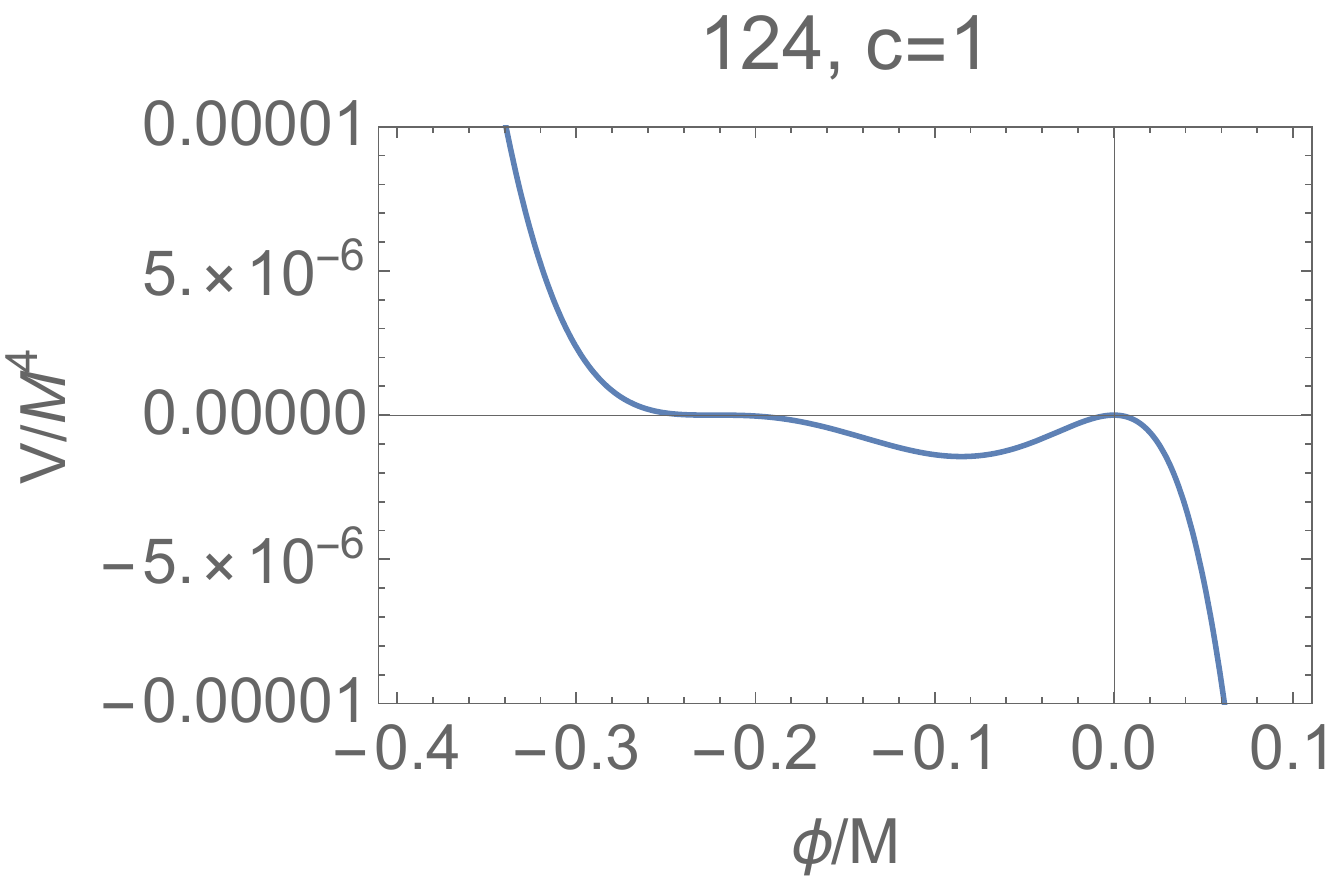}
\includegraphics[width=5cm]{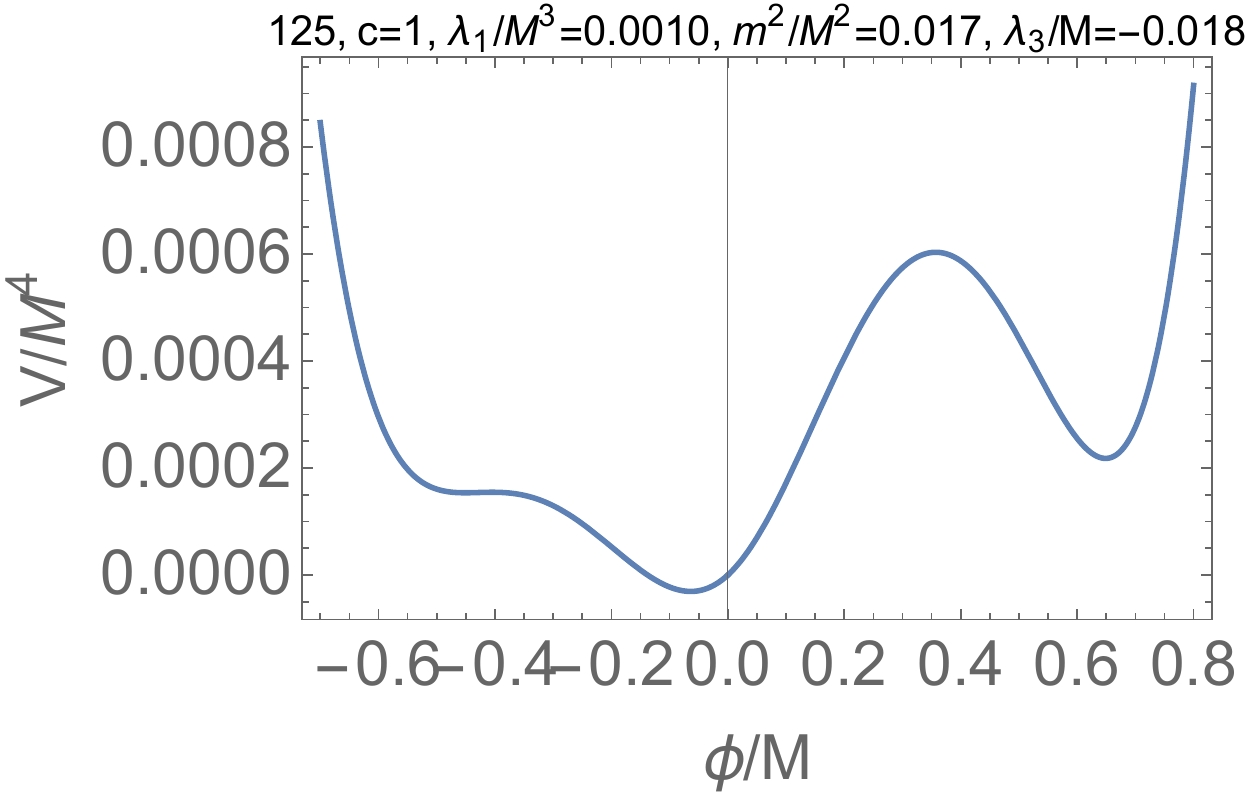}
\includegraphics[width=5cm]{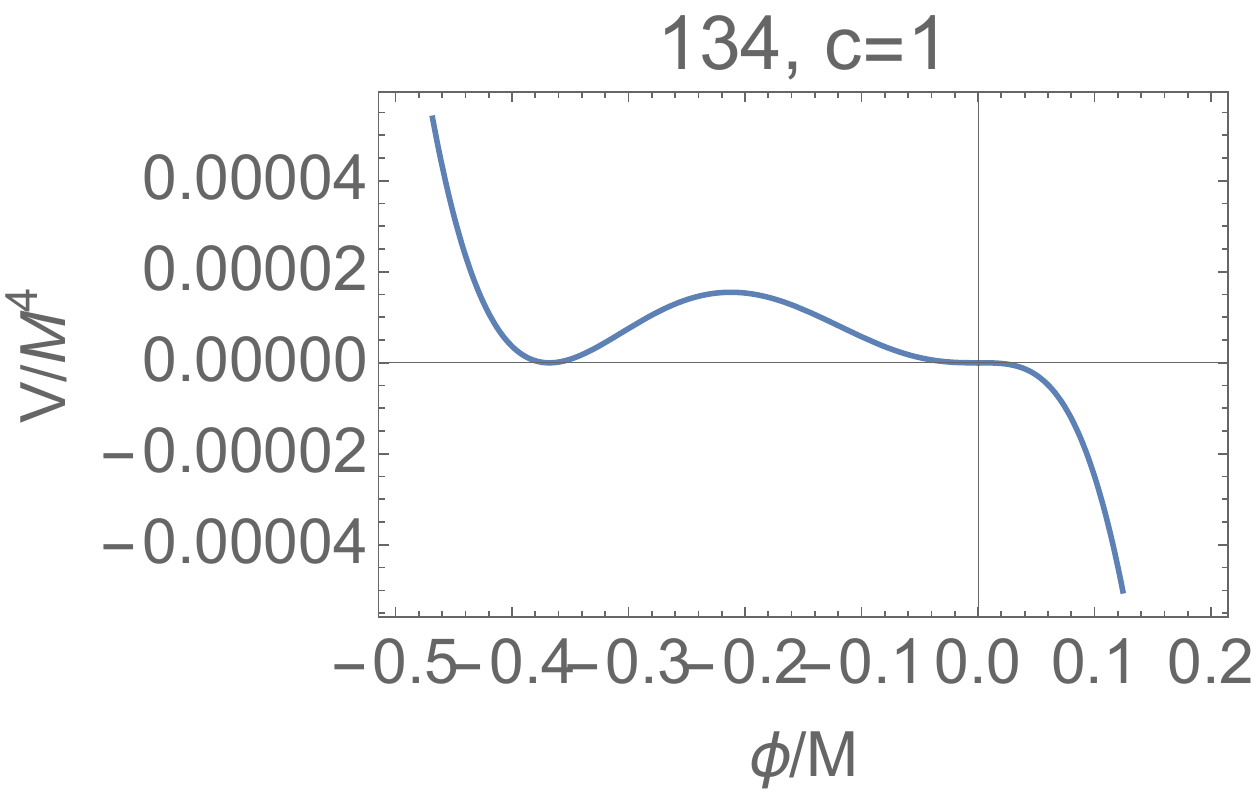}
\includegraphics[width=5cm]{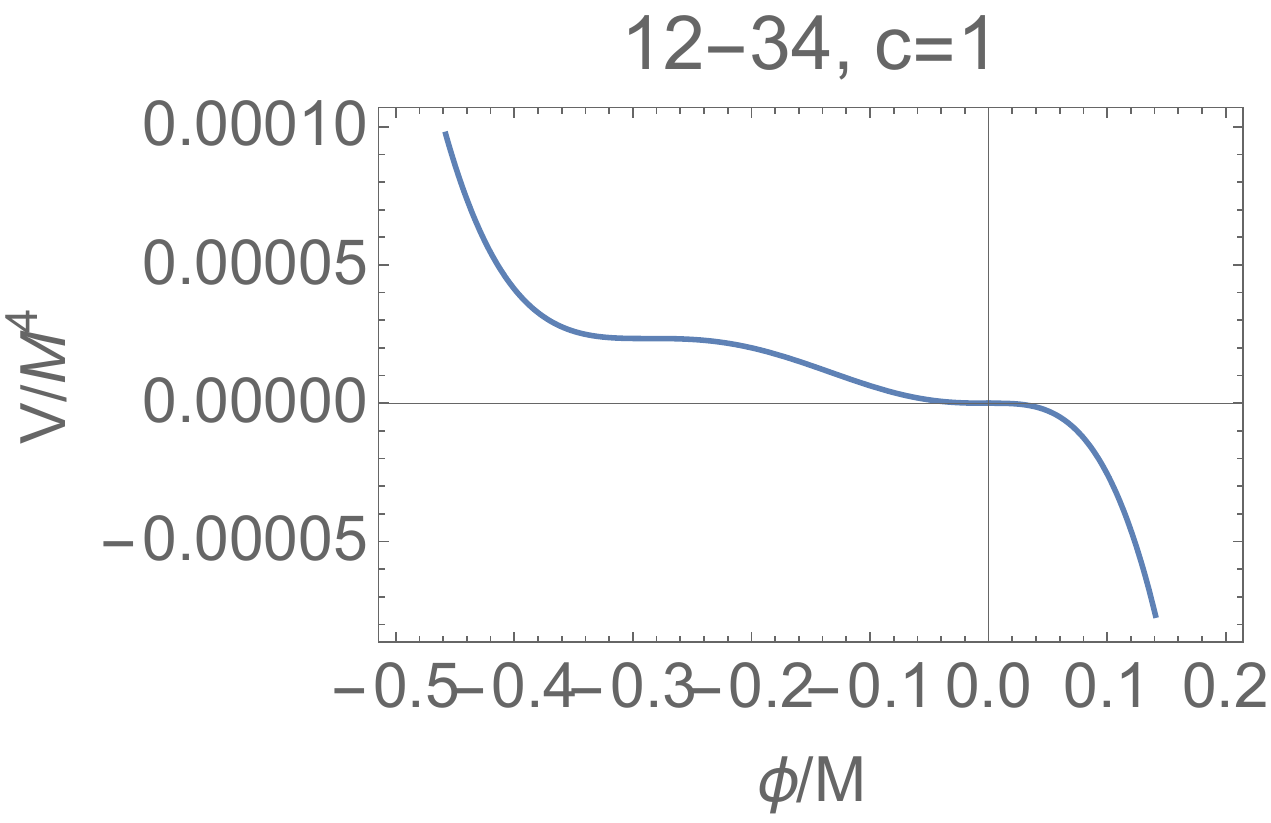}
\includegraphics[width=5cm]{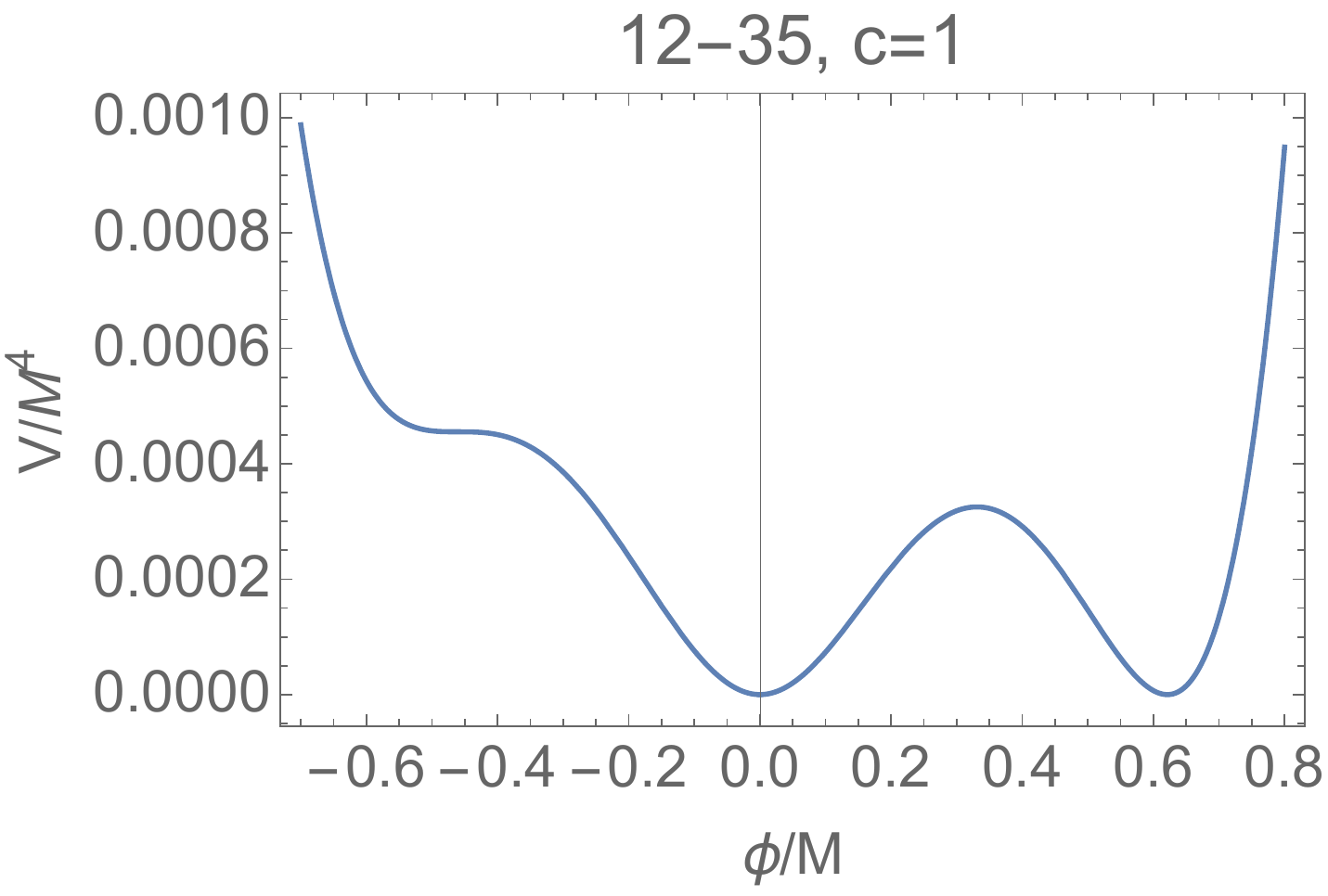}
\includegraphics[width=5cm]{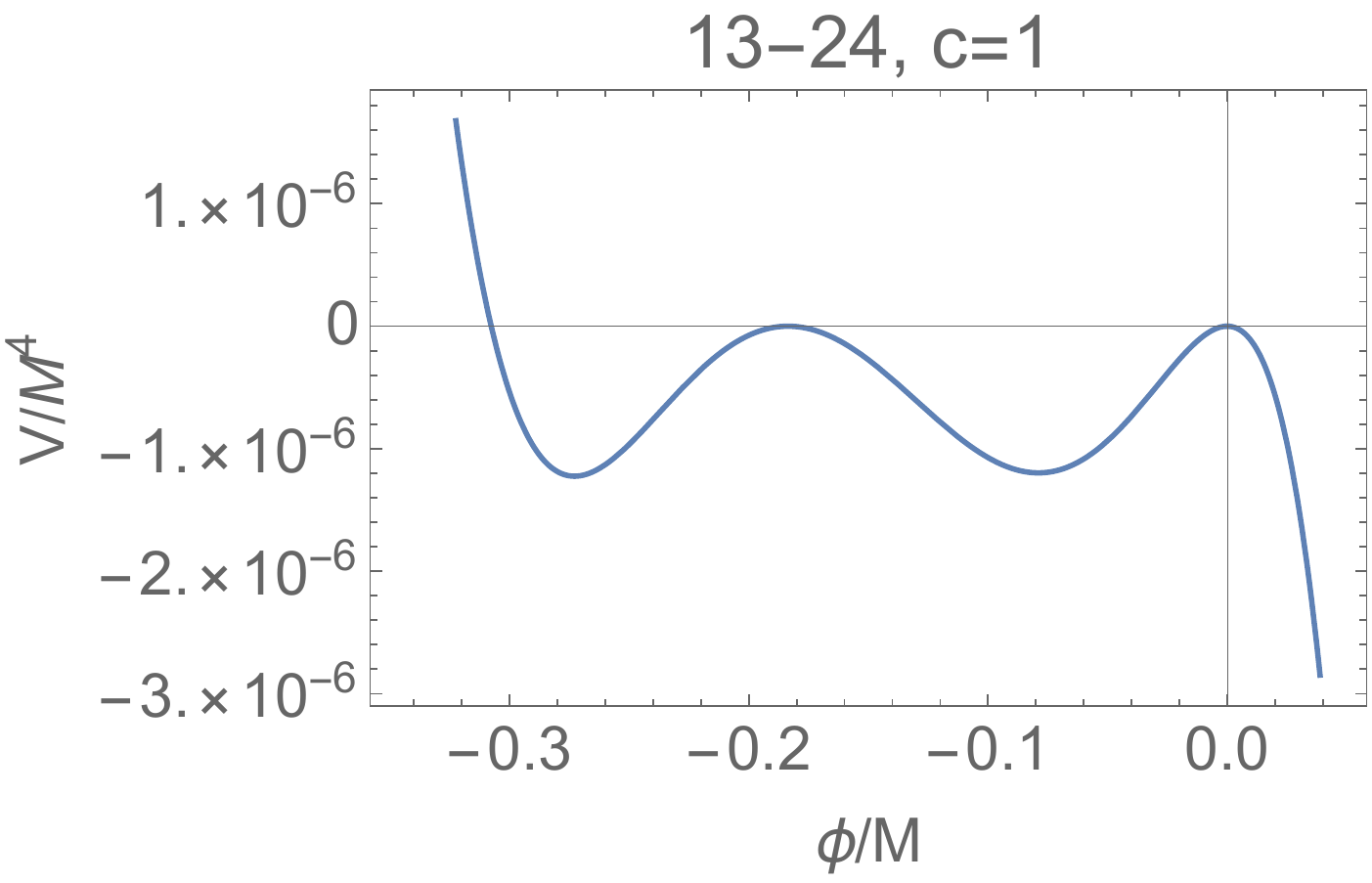}
\includegraphics[width=5cm]{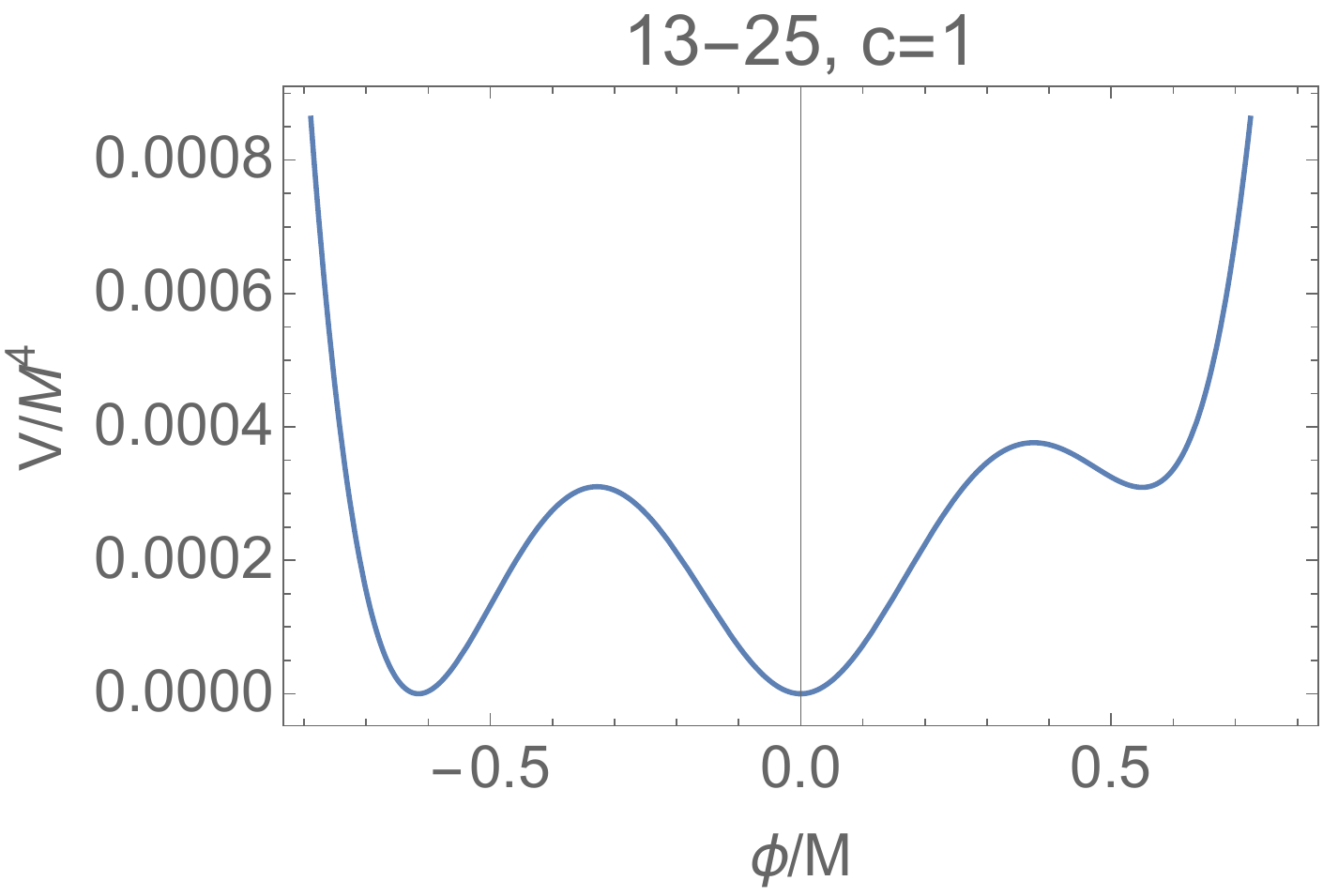}
\includegraphics[width=5cm]{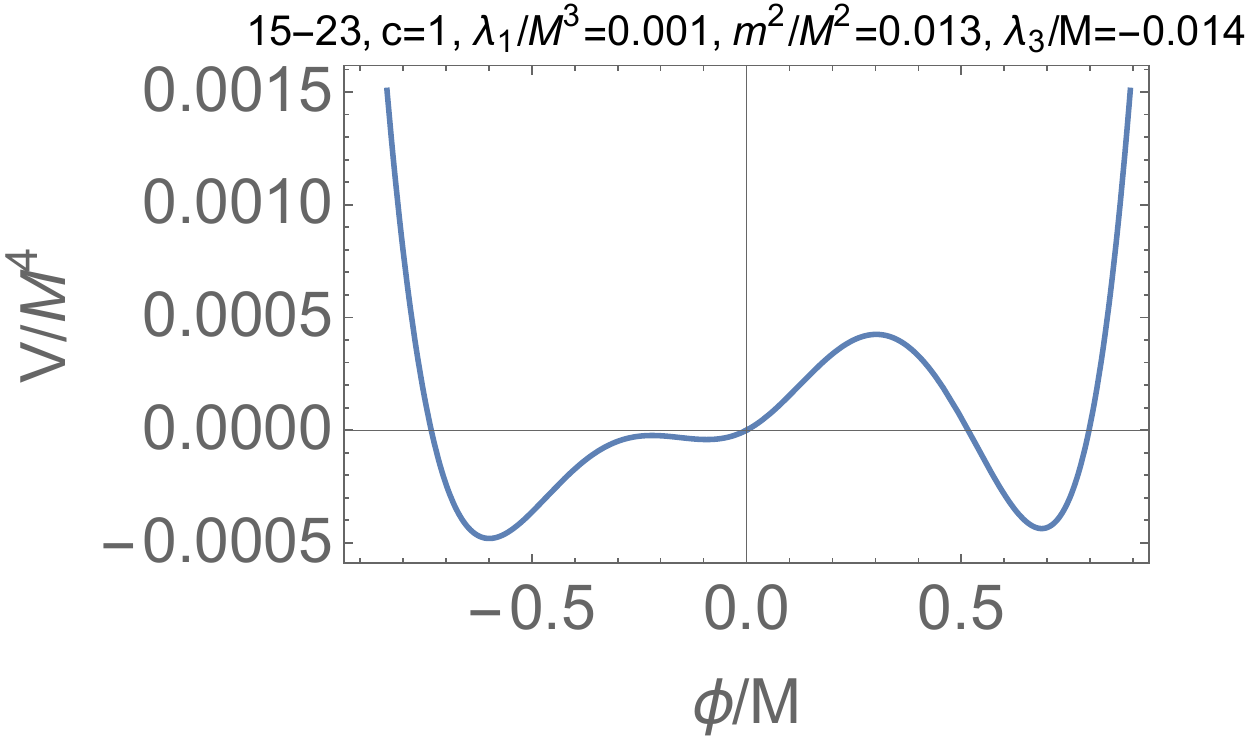}
\caption{The effective potentials corresponding to the bicritical points.  
%
%Each of them corresponds to a multi-critical point in the $(m^2,\lambda_3^{})$ plane. 
%
%Here $c$ is chosen to be $1$.  
} 
\label{fig:bicritical}
\end{center}
\end{figure}
%
%Moreover, most of the above degeneracies can be archived even when $\lambda_1^{}=0$, so we will simply put $\lambda_1^{}=0$ except when it is necessary. 
%
Because these degeneracies are achieved by two-parameter tunings,
each of them lies on a one-dimensional critical curves in the three dimensional
parameter space of $\lambda_{1}^{}, m^2, \lambda_{3}^{}$.
To specify such curves, we consider their intersections with the plane
$\lambda_1^{}= 0$. 
In other word, we will simply set $\lambda_1^{}=0$ except when necessary. 
It will be found that only $125$ and $15-23$ require $\lambda_1^{}\neq 0$. 

\

\noindent $\ast$ {\bf 123}: This critical point is defined by $V'|_{\phi=\phi_S^{}}^{}=V''|_{\phi=\phi_S^{}}^{}=V'''|_{\phi=\phi_S^{}}^{}=0$, and we can solve them analytically by using Eqs.~(\ref{first derivative})(\ref{second derivative})(\ref{third derivative}).   
%
%Thus, in addition to Eq.~(\ref{4-5 line}), we have
%\aln{
%V'''|_{\phi=\phi_S^{}}=\lambda_3^{}+\frac{c}{2}\phi_S^{}\left[\ln\left(\frac{\phi_S^{2}}{M^2e^{1/2}}\right)+\frac{13}{6}\right]=0~. 
%}
The results are  
\aln{\phi_S^{}=\mp Me^{-7/4}~,\quad m^2=-\frac{c}{12}e^{-7/2}M^2~,\quad \lambda_3^{}=\mp \frac{2c}{3}e^{-7/4}M~. 
}
$v_\phi^{}$ and $m_\phi^2$ are numerically given by
\aln{v_\phi^{}\simeq \pm 1.1M~,\quad m_\phi^2=0.26cM^2~. 
} 
The resultant potential is shown in the upper-left panel in Fig.~\ref{fig:bicritical} where we focus on the field region $\phi<0$ so that the degeneracy of the extrema can be easily seen.  

\

\noindent $\ast$ {\bf 124}: This critical point is defined by  $V|_{\phi=\phi_S^{}}^{}=V'|_{\phi=\phi_S^{}}^{}=V''|_{\phi=\phi_S^{}}^{}=0$, and we can also solve them analytically by using Eqs.~(\ref{first derivative})(\ref{second derivative}).  
%This critical point corresponds to the intersecting point between $2-4$, $2-5$, and $4-5$ lines. 
%
%The $4-5$ degeneracy was already solved as Eq.~(\ref{4-5 line}), and \textcircled{\scriptsize 9} is obtained by solving $V|_{\phi=\phi_S^{}}^{}=0$:
%
The results are 
\aln{
\phi_S^{}=\mp Me^{-3/2}~,\quad m^2=-\frac{c}{24}e^{-3}M^2~,\quad \lambda_3^{}=\mp \frac{c}{2}e^{-3/2}M~. 
}
Then, $v_\phi^{}$ and $m_\phi^2$ are numerically solved as 
\aln{v_\phi^{}\simeq \pm 1.1M~,\quad m_\phi^2=0.26cM^2~. 
} 
The resultant potential is shown in the upper-middle panel in Fig.~\ref{fig:bicritical}. 

\

\noindent $\ast$ {\bf 125}: As mentioned before, this critical point needs $\lambda_1^{}\neq 0$ and we have one-parameter solution. 
%
%By the degeneracy $12$, $m^2$ and $\lambda_3^{}$ can be solved as functions of $\phi_S^{}$, and we can numerically find $125$ by putting them into $V(\phi)$.   
%
In the upper-right panel in Fig.~\ref{fig:bicritical}, we show one example of the potential whose parameters are numerically obtained as
%
%Here, the parameters are chosen to be
\aln{
c=1~,\ \lambda_1^{}=0.0010M^3~,\ m^2=0.017M^2~,\ \lambda_3^{}=-0.018M~. 
}
%The results are
%\aln{\phi_S^{}=\mp Me^{-5/4}~,\quad m^2=-\frac{c}{24}e^{-5/2}M^2~,\quad \lambda_3^{}=\mp \frac{c}{2}e^{-5/4}M~.}

\

\noindent $\ast$ {\bf 134}: This critical point can be obtained by putting $m^{2}=0$ and solving $V|_{\phi=v_f^{}}^{}=V'|_{\phi=v_f^{}}^{}=0$ with respect to $\lambda_3^{}$ and $v_f^{}$. 
%
%Note that the resultant solution of $\phi$ corresponds to $v_f^{}$, not to $v_\phi^{}$.    
% 
The solutions are 
\aln{
\lambda_3^{}=\pm \frac{c}{4}Me^{-1}~,\quad v_f^{}=\pm Me^{-1}~. 
}
Then, the true vacuum is determined by 
\aln{\frac{\partial V}{\partial \phi}\bigg|_{\phi=v_\phi^{}}^{}=\frac{cv_\phi^2}{8}\left(v_f^{}+\frac{2}{3}v_\phi^{}\log\left(\frac{v_\phi^2}{M^2e^{-1/2}}\right)\right)=0~\quad \therefore\  3v_f^{}+4v_\phi^{}\log\left(\frac{|v_\phi^{}|}{Me^{-1/4}}\right)=0~,
}
whose solution is 
\aln{
v_\phi^{}=\mp M\exp\left[W\left(\frac{3|v_f^{}|}{4Me^{-1/4}}\right)-\frac{1}{4}\right]=\mp M\exp\left[W\left(\frac{3}{4e^{3/4}}\right)-\frac{1}{4}\right]=\mp 1.0M~, 
}
where $W(z)$ is the Lambert $W$ function satisfying $z=e^WW$. 
The mass of $\phi$ is 
\aln{m_\phi^2=\frac{\partial V^2}{\partial \phi^2}\bigg|_{\phi=v_\phi^{}}=\frac{c}{6}\left(1+W\left(\frac{3}{4e^{3/4}}\right)\right)v_\phi^2=0.22 cM^2~. 
}
The resultant potential is shown in the middle-left panel in Fig.~\ref{fig:bicritical}.  

\

\noindent $\ast$ {\bf 135 (15-24)}: This critical point is the  same as $135-24$.   
Thus, the parameters are given by Eq.~(\ref{parameters in 135}). 
%can be obtained by putting $\lambda_3^{}=0$ and solving $V|_{\phi=v_\phi^{}}^{}=V'|_{\phi=v_\phi^{}}^{}=0$ as functions of $m^2$ and $v_\phi^{}$.  
%
%The results are 
%\aln{
%m^2=\frac{c}{24}M^2e^{-1/2}~,\quad v_\phi^{}=\pm Me^{-1/4}~,\quad  m_\phi^2=%\frac{\partial V^2}{\partial \phi^2}\bigg|_{\phi=v_\phi^{}}=
%\frac{c}{12}M^2e^{-1/2}~.
%}
%We show the resultant potential in the middle panel on the second row in Fig.~\ref{fig:bicritical}. 

\

\noindent $\ast$ {\bf 234}: This critical point is also the  same as $234-15$.    
Thus, the potential and the parameters are given by Eqs.~(\ref{potential 234-15})(\ref{mphi 234-15}).
%This critical point is a trivial one, $m^2=\lambda_3^{}=0$, and the potential is simply given by
%\aln{
%V(\phi)=\frac{c}{2\cdot 4!}\phi^4\log\left(\frac{\phi^2}{M^2}\right)~,}
%which has a minimum at $v_\phi^{}=\pm Me^{-1/4}$. 
%
%The mass of $\phi$ is
%\aln{m_\phi^{2}=\frac{c}{6}M^2e^{-1/2}~.}
%The resultant potential is shown in the right panel on the second row in Fig.~\ref{fig:bicritical}. 

\

\noindent $\ast$ {\bf 12-34}: By the saddle point conditions $V'|_{\phi=\phi_S^{}}^{}=V''|_{\phi=\phi_S^{}}^{}=0$, $m^2$ and $\lambda_3^{}$ can be solved as functions of $\phi_S^{}$:    
\aln{
m^2=\frac{c\phi_S^2}{12}\left(\frac{5}{2}+\log\left(\frac{\phi_S^{2}}{M^2}\right)\right)~,\quad \lambda_3^{}=-\frac{c\phi_S^{}}{3}\left(\frac{3}{2}+\log\left(\frac{\phi_S^{2}}{M^2}\right)\right)~. \label{4-5 line}
}
%
%For example, when we put $\phi_S^{}=Me^{-1/2}$ which corresponds to \textcircled{\scriptsize 4}, we reproduce the mass parameter in Eq.~(\ref{vacuum 4}). 
%
The critical point $12-34$ is given by the solution of $m^2=0$;
\aln{\lambda_3^{}=\pm \frac{c}{3}Me^{-5/4}~,\quad \phi_S^{}=\pm Me^{-5/4}~, 
}
and the corresponding true vacuum is determined by 
%ãšãªãã?ãã®æã?true vacuumã¯$V'$ã«äžã?$\Lambda$ãä»£å¥ããŠ($\phi<0$ã®solutionã§ããããšã«æ³šæããŠ)
\aln{
&\frac{\partial V}{\partial \phi}\bigg|_{\phi=v_s^{}}^{}=\frac{c}{6}v_\phi^2\left(\phi_S^{}+\frac{v_\phi^{}}{2}\log\left(\frac{v_\phi^2}{M^2e^{-1/2}}\right) \right)=0~, %\tilde{\phi}_S^{}=-\tilde{v}_\phi^{}\log(-\tilde{v}_\phi^{})
\\
\therefore\ v_\phi^{}&=\mp M\exp\left[W\left(\frac{|\phi_S^{}|}{Me^{-1/4}}\right)-\frac{1}{4}\right]=\mp M \exp\left[W\left(e^{-1}\right)-\frac{1}{4}\right]=\mp 1.0M
%=-\phi_S^{}e^{1+W(e^{-1})}=-\phi_S^{}W^{-1}(e^{-1})~.%\quad e^{W(e^{-1})}=1.32
}
The mass of $\phi$ is given by
\aln{m_\phi^2 =\frac{\partial V^2}{\partial \phi^2}\bigg|_{\phi=v_\phi^{}}=\frac{c}{6}\left(1+W(e^{-1})\right)v_\phi^2=0.23cM^2~. %\quad \therefore\ C=1+W(e^{-1}) 
}
The resultant potential is shown in the center panel in Fig.~\ref{fig:bicritical}. 

\

\noindent $\ast$ {\bf 12-35}: This critical point can be obtained by putting Eq.~(\ref{4-5 line}) into the potential and solving $V|_{\phi=v_\phi^{}}^{}=V'|_{\phi=v_\phi^{}}^{}=0$ as functions of $v_\phi^{}$ and $\phi_S^{}$. 
%
%Those equations are respectively given by 
%\begin{numcases}{}
%24\phi_S^2-16\phi_S^{}v_\phi^{}-3v_\phi^2+4\phi_S^{}(3\phi_S^{}-4v_\phi^{})\log\left(\frac{\phi_S^{2}}{M^2}\right)+6v_\phi^2\log\left(\frac{v_\phi^{2}}{M^2}\right)^2=0~, 
%\\
%2\phi_S^2-2\phi_S^{}v_\phi^{}+\phi_S^{}(\phi_S^{}-2v_\phi^{})\log\left(\frac{\phi_S^{}}{M}\right)^2+v_\phi^2\log\left(\frac{v_\phi^{2}}{M^2}\right)=0~. 
%\end{numcases}
%
The numerical results are 
\aln{v_\phi^{}\simeq \pm 0.62M~,\quad \phi_S^{}\simeq \mp 0.46 M~,\quad m^2\simeq 0.017 cM^2~,\quad \lambda_3^{}\simeq  \mp  0.0075 c M~. 
}
Then, the mass of $\phi$ is calculated as 
\aln{
m_\phi^2=\frac{\partial^2 V}{\partial \phi^2}\bigg|_{\phi=v_\phi^{}}\simeq\begin{cases}
   0.017cM^2  & \text{for $v_\phi^{}=0$}
\\
 0.033 c M^2  & \text{for $|v_\phi^{}|=0.62M$} 
 \end{cases}~.
 %\simeq 0.085cv_\phi^2:=\frac{\lambda_{\phi S}^2}{32\pi^2} Cv_\phi^2\quad \therefore\ C=0.085\times 2=0.17~, 
}
The resultant potential is shown in the middle-right panel in Fig.~\ref{fig:bicritical}.  

\

\noindent $\ast$ {\bf 12-45}: This critical point is the same as $1245$. 
Thus, the parameters are given by Eqs.~(\ref{sol1 1245})(\ref{sol2 1245}). 
%corresponds to the degeneracies between  four vacua, $1-2-4-5$. 
%
%By putting $\lambda_3^{}=0$ and solving $V'|_{\phi=\phi_S^{}}^{}=V''|_{\phi=\phi_S^{}}^{}=0$ as functions of $\phi_S^{}$ and $m^2$, we obtain 
%\aln{
%m^2=\frac{c}{12e}M^2,\quad \phi_S^{}=Me^{-1/2}~,  \label{vacuum 4}
%}
%from which $m_\phi^2$ is calculated as
%\aln{m_\phi^2=\frac{\partial V^2}{\partial \phi^2}\bigg|_{\phi=0}^{}=\frac{c}{12e}M^2~. } 
%The resultant potential is shown in the upper-right panel in Fig.~\ref{fig:vacuum}. 
%
%This case is phenomenologically less attractive because we can not realize symmetry breaking by $\phi$.   
%
%

\

\noindent $\ast$ {\bf 13-24}: %This critical point corresponds to the intersecting point between $2-4$ and $3-5$ lines. 
%
%The solution of $24$ is the same as Eq.~(\ref{condition_mpp 6}) but $v_\phi^{}$ is replaced by the local maximum $\phi=v_m^{}$.   
The $24$ degeneracy is defined by $V|_{\phi=\phi_2^{}}^{}=V'|_{\phi=\phi_2^{}}^{}=0$ and this can be solved as 
\aln{
m^2=\frac{c}{24}\phi_2^2\left(2+1\log\left(\frac{\phi_2^{2}}{M^2}\right)\right)~,\quad \lambda_3^{}=-\frac{c}{4}\phi_2^{}\left(1+\log\left(\frac{\phi_2^{2}}{M^2}\right)\right)~,%\quad v_m^{}>0~. 
\label{condition 13-24}
}
Then, the degeneracy between $1$ and $3$ can be numerically solved by putting Eq.~(\ref{condition 13-24}) into $V(\phi)$ and changing $\phi_2^{}$.  
The results are 
\aln{%v_m^{}\simeq 0.24M~,\quad 
v_\phi^{}\simeq \pm 1.1M~,\quad m^2\simeq-0.0020c M^2~,\quad \lambda_3^{}\simeq\mp 0.11cM~,\quad m_\phi^2\simeq 0.25cM^2~.%=\frac{\lambda_{\phi S}^2}{32\pi^2}Cv_\phi^2,\quad \therefore\ C= 0.22\times 2=0.44,
}
The resultant potential is shown in the lower-left panel in Fig.~\ref{fig:bicritical}. 

\

\noindent $\ast$ {\bf 13-25}: %This critical point corresponds to the intersecting point between $1-3$ and $2-5$ lines. 
The $13$ degeneracy is the same as Eq.~(\ref{condition 13-24}) with the replacement $\phi_2^{}\rightarrow \phi_1^{}$. 
Then, the degeneracy $25$ can be numerically found by putting them into $V(\phi)$ and changing $\phi_1^{}$.   
The results are 
\aln{%v_\phi^{}\simeq \pm 0.79 M~,\quad
 m^2\simeq 0.016cM^2~,\quad \lambda_3^{}\simeq \mp 0.0046cM~,\quad 
(v_\phi^{},m_\phi^{2})\simeq \begin{cases} (\pm 0.62M,0.032cM^2)  & {\rm  for}\ 1 
\\
(0,0.016cM^2)  & {\rm  for}\ 3
\end{cases}~.
%:=\frac{\lambda_{\phi S}^2}{32\pi^2}Cv_\phi^2,\quad \therefore\  C=0.085\times 2=0.17
}
The resultant potential is shown in the lower-middle panel in Fig.~\ref{fig:bicritical}.  

\

\noindent $\ast$ {\bf 15-23}: As mentioned before, this critical point needs $\lambda_1^{}\neq 0$ and we have one-parameter solution. 
%
%By the degeneracy $12$, $m^2$ and $\lambda_3^{}$ can be solved as functions of $\phi_S^{}$, and we can numerically find $125$ by putting them into $V(\phi)$.   
%
In the lower-right panel in Fig.~\ref{fig:bicritical}, we show one example of the potential whose parameters are numerically obtained as
%
%Here, the parameters are chosen to be  
\aln{
c=1~,\ \lambda_1^{}=0.0010M^3~,\ m^2=0.013M^2~,\ \lambda_3^{}=-0.014M~. 
}
%The results are
%\aln{\phi_S^{}=\mp Me^{-5/4}~,\quad m^2=-\frac{c}{24}e^{-5/2}M^2~,\quad \lambda_3^{}=\mp \frac{c}{2}e^{-5/4}M~. }
In Table.~\ref{tab:case}, we summarize all the doubly critical points corresponding to 2-parameter tuning.  
%
%Here, ``{\it Degeneracy}" means the type-A criticality,  i.e. $V|_{\phi=\phi_i^{}}^{}=V|_{\phi=\phi_j^{}}^{}~(i\neq j)$. 
%
%
\begin{table}[!t]
\begin{center}
\begin{tabular}{|c||cccc|}\hline
            &  $\mathbb{Z}_2 $  & $|v_\phi^{}|$    & $m_\phi^2$      & $\lambda_1^{}$  %& Symmetry breaking 
            \\\hline\hline
{\bf 123} &  Broken     & $1.1M$     &  $0.26cM^2$  &  0
 \\\hline
{\bf 124}  &  Broken    & $1.1M$         &  $0.26cM^2$  & 0
 \\\hline
{\bf 125}  &  Broken  & -  &   -         & $\neq 0$
 \\\hline
{\bf 134} &  Broken  & $Me^{-1/4}\exp\left\{W(3e^{-3/4}/4))\right\}$   &  $cv_\phi^2(1+W(3e^{-3/4}/4))/6$          & 0
 \\\hline
{\bf 135 (15-24)} & Exact   & $Me^{-1/2}$   
 &  $cM^2e^{-1}/12$          & 0 
   \\\hline
{\bf 234} &  Exact  & $Me^{-1/4}$  &     $cM^2e^{-1/2}/6$       & 0
    \\\hline
{\bf 12-34} &  Broken  & $Me^{-1/4}\exp\left\{W(e^{-1}))\right\}$    &  $cv_\phi^2(1+W(e^{-1}))/6$           & 0
    \\\hline
{\bf 12-35} &  Broken  & $0$ or $0.62M$    & $0.017cM^2$ or $0.033cM^2$            & 0 
   \\\hline
{\bf 12-45} & Exact   & $0$   &  $cM^2e^{-3/2}/12$           & 0 
    \\\hline
{\bf 13-24} &  Broken  & $1.1M$  &  $0.25cM^2$            &  0
    \\\hline
 {\bf 13-25} &  Broken  & $0$ or $0.62M$  &  $0.016cM^2$ or $0.032cM^2$           &  0
    \\\hline
  {\bf 15-23} &  Broken  & -  &  -            &  $\neq 0$
    \\\hline
\end{tabular}
\caption{
All the critical points of the one-loop effective potential Eq.~(\ref{CW potential 3}) corresponding to   2-parameter tuning.     
}
\label{tab:case}
\end{center}
\end{table}
%
%One can see that phenomenological predictions of \textcircled{\scriptsize 6} and  \textcircled{\scriptsize 7} (\textcircled{\scriptsize 8}, \textcircled{\scriptsize 9}, and \textcircled{\scriptsize 10}) are almost identical each other because they are related through the tiny changes of parameters.  
% 

%
\bibliography{Bibliography}
\bibliographystyle{utphys}

\end{document}